\documentclass[fleqn,usenatbib,useAMS]{mnras}

\usepackage{graphicx}	
\usepackage{amsmath}	
\usepackage{amssymb}	
\usepackage{multicol}        
\usepackage{bm}		
\usepackage{pdflscape}	
\usepackage{xcolor}
\usepackage{url}

\usepackage{newtxtext,newtxmath}

\usepackage[T1]{fontenc}

\DeclareRobustCommand{\VAN}[3]{#2}
\let\VANthebibliography\thebibliography
\def\thebibliography{\DeclareRobustCommand{\VAN}[3]{##3}\VANthebibliography}

\newcommand{\se}{{\sc SExtractor\ }}
\newcommand{\psf}{{\sc PSFEx\ }}
\newcommand{\dos}{{\sc SExtractor+PSFEx\ }}

\usepackage{graphicx}	
\usepackage{amsmath}	
\usepackage{amssymb}	
\usepackage{multicol}        
\usepackage{longtable}
\usepackage{lscape} 
\usepackage{bm}		
\usepackage{pdflscape}	
\usepackage[flushleft]{threeparttable}

\title[Galaxy clustering in the VVV Near-IR Galaxy Catalogue] 
{Galaxy clustering in the VVV Near-IR Galaxy Catalogue}

\author[Soto et al.]
{
\parbox[t]{\textwidth}{Mario Soto$^{1}$, Mario A. Sgr\'o$^{2,3}$, Laura D. Baravalle$^{2,3}$, M. Victoria Alonso$^{2,3}$, Jos\'e Luis Nilo Castell\'on$^{4,5}$, Carlos Valotto$^{2,3}$, 
 Antonela  Taverna$^{2,3}$, 
Eugenia D\'iaz-Gim\'enez$^{2,3}$,
Carolina Villal\'on$^{3}$, Dante Minniti$^{6,7,8}$}
\vspace*{6pt} \
\\
$^{1}$ Instituto de Astronom\'ia y Ciencias Planetarias, Universidad de Atacama, Copayapu 485, Copiap\'o, Chile.\\
$^{2}$ Observatorio Astron\'omico de C\'ordoba, Universidad Nacional de C\'ordoba, Laprida 854, X5000BGR, C\'ordoba, Argentina.\\
$^{3}$ Instituto de Astronom\'ia Te\'orica y Experimental (IATE, CONICET-UNC). C\'ordoba, Argentina.\\
$^{4}$ Direcci\'on de Investigaci\'on y Desarrollo, Universidad de La Serena. Av. Ra\'ul Bitr\'an Nachary 1305, La Serena, Chile.\\
$^{5}$ Departamento de Astronom\'ia, Universidad de La Serena. Av. Juan Cisternas 1200, La Serena, Chile.\\
$^{6}$ Departamento de F\'isica, Facultad de Ciencias Exactas, Universidad Andr\'es Bello, Av. Fernandez Concha 700, Las Condes, Santiago, Chile.\\
$^{7}$ Instituto Milenio de Astrof\'isica, Santiago, Chile.\\
$^{8}$ Vatican Observatory, V00120 Vatican City State, Italy.\\
}

\begin{document}

\date{}

\pagerange{\pageref{firstpage}--\pageref{lastpage}} \pubyear{2010}

\maketitle

\label{firstpage}

\begin{abstract}

Mapping galaxies at low Galactic latitudes and determining their clustering status are fundamental steps in defining the large-scale structure in the nearby Universe. 
The VVV Near-IR Galaxy Catalogue (VVV NIRGC) allows us to explore this region in great detail.
Our goal is to identify 
galaxy overdensities and characterize galaxy clustering in the Zone of Avoidance. 
We use different clustering algorithms to identify galaxy overdensities: the Voronoi tessellations, the Minimum Spanning Tree and the Ordering Points To Identify the Clustering Structure.  We studied the membership, isolation, compactness, and flux limits to identify compact groups of galaxies.  Each method identified a variety of galaxy systems across the Galactic Plane that are publicly available.
We also explore the probability that these systems are formed by concordant galaxies using mock catalogues.  
Nineteen galaxy systems were identified in all of the four methods.
They have the highest probability to be real overdensities.  
We stress the need for spectroscopic follow-up observations to confirm and characterize these new structures.

\end{abstract}


\section{Introduction}

Most of the galaxies in the Universe inhabit vast cosmological structures, such as groups, clusters and filaments \citep{Peebles1980}.
The conjunction of all these structures together with other giant and extreme low-density areas called ``voids'' conforms the Large-Scale Structure of the Universe (LSS, \citealt{Bond1996}).  This structure is studied using both observations and simulations  over the whole sky with the exception of those regions in which the Milky Way (MW) is placed. In this hidden part of the sky, it is difficult to observe galaxies that lie behind the MW plane \citep{Kraan2018} due to the high stellar density and the high level of Galactic extinction. This region is called the Zone of Avoidance (ZoA) and the first studies of background galaxies were made mainly in the optical and radio wavelengths \citep{Kraan1999,Kraan2000}. 
The observations at near-infrared (NIR) wavelengths are weakly sensitive to Galactic extinction and allow to search for galaxies behind the regions blocked by the MW.  The pioneer work on this waveband was the Two Micron All Sky Survey (2MASS; \citealt{Skrutskie2006}), which observed  the 99.998\% of the celestial sphere in the $J$, $H$ and $K_{s}$ NIR passbands. \cite{2000AJ....119.2498J} developed an algorithm capable of identifying extended sources or candidate galaxies from the 2MASS survey.
The 2MASS extended source catalogue arose as a suitable catalogue for identifying systems of galaxies not only because of its full-sky coverage, but also because the $K_s$-band photometry is a very good tracer of the stellar mass content of galaxies. The first catalogue of clusters in the near Universe built from the early releases of 2MASS was presented by \cite{Kochanek+03}. 
When redshift information became available mainly via the 2MASS Redshift survey (2MRS, \citealt{Huchra2005,Huchra2012}) and the 2M++ galaxy redshift compilation (\citealt{lavaux+11}), catalogues of galaxy systems from these sources were produced by several authors using different methods and algorithms
\citep{crook+07,lavaux+11,DiazGimenez+12,tully15,DiazGimenez+15,tempel+16,Lim+17,Kourkchi2017}.
Recently, \cite{Lambert2020} used an improved friends-of-friends algorithm \citep{Huchra1982} to generate a galaxy group catalogue including some galaxies in the ZoA based on an updated version of the 2MRS \citep{Macri2019}.  However, none of these works has covered the inner parts of the ZoA (Galactic latitude  $|b|<5^\circ$). 

The VISTA Variables in the Milky Way (VVV, \citealt{Minniti2010} is a deeper NIR survey and it has become a new tool to search for galaxies in the ZoA behind the MW. \cite{Amores2012} carried out the first work applying a combination of colour cuts and visual inspection, finding more than 200 galaxy candidates in the d003 tile (covering only 1.5 sq. deg.) behind the MW disc. \cite{Coldwell2014} studied the NIR properties of the galaxy cluster Suzaku J1759-3450 \citep{Mori2013}, located behind the Galactic bulge and discovered by the Suzaku \citep{Mitsuda2007} and Chandra \citep{Weisskopf2000} X-ray space telescopes. 
Recently, \cite{Galdeano2021} used the VVV data and mock catalogues to analyse the b204 tile that corresponds to the Galactic bulge, reporting 607 new galaxy candidates, of which only 17 had previous identifications. 
Our main goal is to identify and characterize extragalactic sources on the regions behind the Southern disc of the MW. We developed a photometric procedure  on the images of the VVV survey to that end.  In \cite{Baravalle2018}, we established the basic methodology to search for galaxies in the reddened and crowded fields at low Galactic latitudes. This method was tested in the d010 and d115 tiles of the disc, and revealed 530 new galaxy candidates. In \cite{Baravalle2019}, we presented the first confirmed galaxy cluster VVV-J144321.06-611753.9 at $z$ = 0.234 found in the VVV d015 tile.  We detected the galaxies using our photometric
procedure and the galaxy cluster was identified using the Cluster Red Sequence in the colour-magnitude diagram \citep{Gladders2000}. The NIR spectra of the two brightest galaxies allowed us to confirm their extragalactic nature with characteristics of early-type galaxies.  
More recently,  we applied this methodology to the 152 VVV tiles of these regions using the $J$, $H$ and $K_{s}$ images \citep{Baravalle2021} creating the VVV NIR galaxy catalogue (VVV NIRGC), which contains a total of 5563 galaxies. 


In this paper we apply different clustering methods to the VVV NIRGC in order to identify galaxy systems as candidates of compact and loose groups in the ZoA. The paper is organized as follows. Section $\S$2  briefly explains the photometric and spectroscopic data used in this work.   Section $\S$3 describes the four different methods involved in the detection of the underlying structures.  
The input parameters used in these methods were tested and chosen taking into account mock catalogues detailed in the Appendix.  
Section $\S$4 contains the main results of our analysis. Finally, section $\S$5 presents the summary and conclusions of this work.

\section {Photometric and spectroscopic data}

The VVV is a NIR variability survey of the entire MW bulge and a large portion of the Southern Galactic disc \citep{Minniti2010}.
In order to detect and characterize extended sources, we applied \dos in the $J$, $H$ and $K_{s}$ images of the 152 VVV tiles of the Southern Galactic disc (\citealt{Baravalle2021} and references therein). 
\se identifies  objects from an astronomical image \citep{Bertin1996} and  \psf extracts models of the best Point Spread Function (PSF) from the image \citep{Bertin2011}.  Then, \se applies the PSF models to each detected source to obtain the astrometric, photometric and morphological properties. 
From these properties, we attempted a first discrimination between extended and point sources. 

These extended sources satisfied the morphological criteria, which are the stellar index \textit{CLASS\_STAR}~$< 0.3$; the SPREAD\_MODEL ($\Phi$) parameter, $\Phi> 0.002$;  the radius that contains 50\% of the total flux of an object ($R_{1/2}$), $1.0 < R_{1/2} < 5.0 \ \rm arcsec$; the concentration index ($C$, \citealt{Conselice2000}), $2.1 < C < 5$ 
and the extinction-corrected colour cuts of 
$0.5<(J - K_{s})^{\circ}< 2.0 \ \rm mag$, 
$0.0 < (J - H)^{\circ}< 1.0 \ \rm mag$,  
$0.0 < (H - K_{s})^{\circ}< 2.0 \ \rm mag$, and 
$(J - H)^{\circ} + 0.9 (H - K_{s})^{\circ} > 0.44 \ \rm mag$. 
With these colour restrictions, we also minimized the number of false detections following  \cite{Baravalle2018} procedure. All candidates that satisfied these criteria were visually inspected using the images in the $J$, $H$ and $K_{s}$ passbands and the false-colour RGB images in case of doubts.  The inspections together with the morphological and colour constraints were used to confirm galaxies.  

\begin{figure*}
\begin{center}
\includegraphics[width=\textwidth]{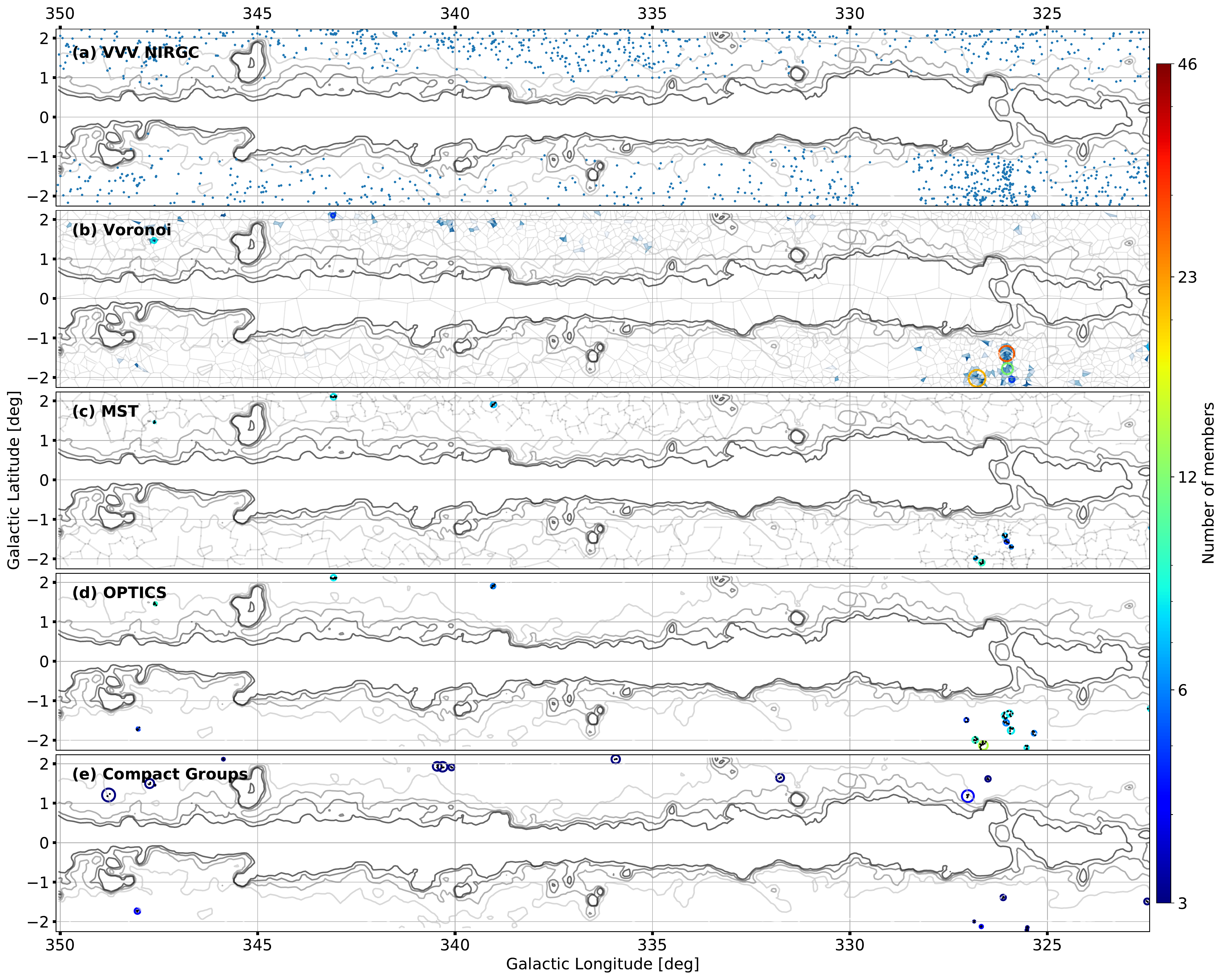}
\end{center}
\caption{Southern Galactic disc: Galactic longitudes between 350$^\circ$ to 322$^\circ$ and 322$^\circ$ to 295$^\circ$. The distribution of galaxies from the VVV NIRGC is shown in panel (a) and the angular distributions of overdensities identified with the four methods in panels (b) to (e).  The colour code is related to the number of members and the circle size represents the angular radius of the minimum circle that encloses all the members. For CGs, the radii have been enhanced by a factor of four for better visualization. The A$_{V}$ iso-contours derived from the extinction maps of \citet{Schlafly2011} are superimposed in a grey gradient with levels of 10, 15, 20 and 25 mag.}
\label{fig:all_1}
\end{figure*}

\begin{figure*}
   \includegraphics[width=\textwidth]{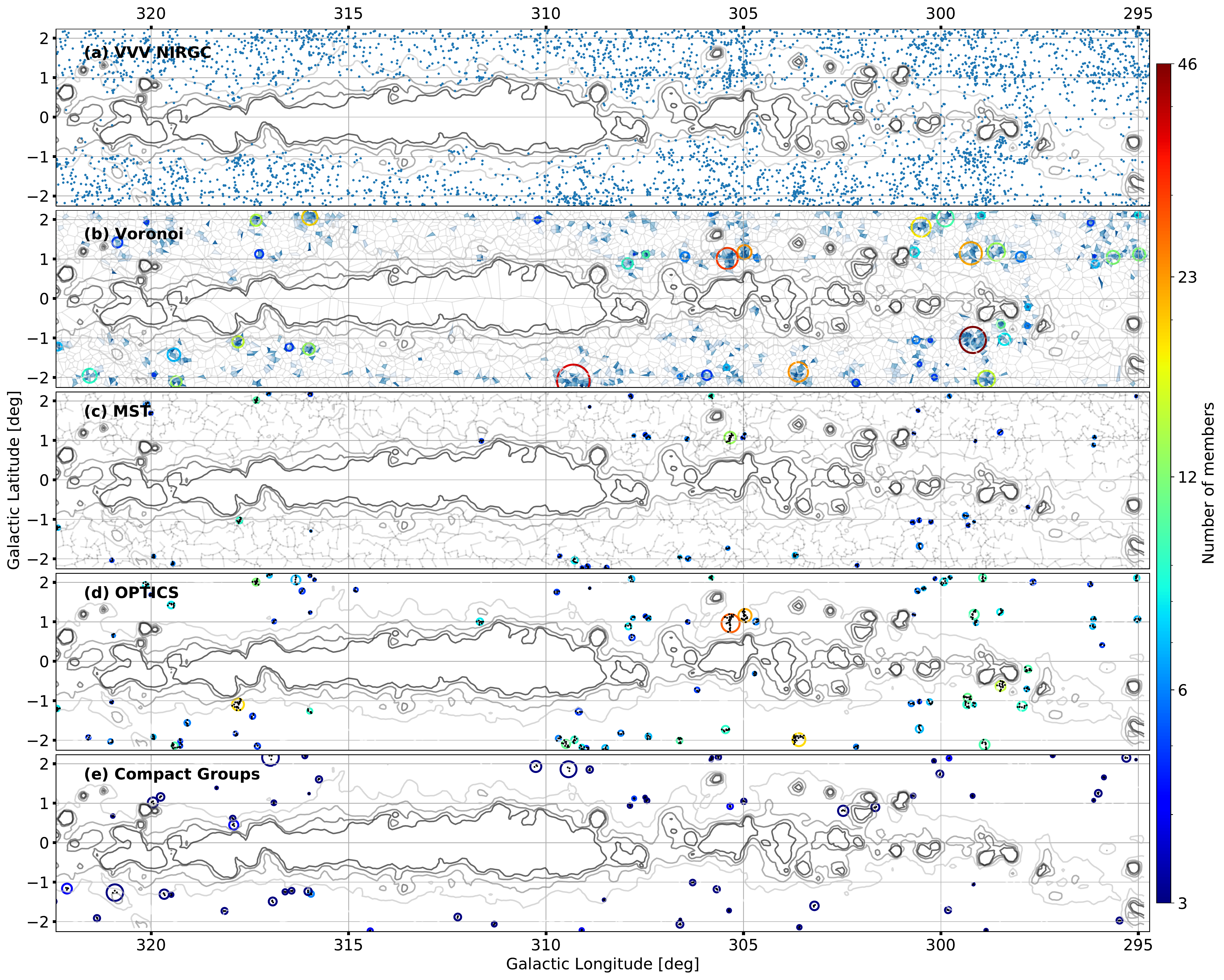}
\contcaption{}
\label{}
\end{figure*}

The VVV NIRGC (\citealt{Baravalle2021}) is the result of our study across the area defined within $295^{\circ}<l< 350^{\circ}$ and
$-2.25^{\circ}< b < +2.25^{\circ}$ in the VVV survey.  The catalogue contains 5563 galaxies in 
the Southern Galactic disc, 99\% of which are new discoveries. 
At these low Galactic latitudes only 45 galaxies were previously observed by other authors:  12 galaxies 
by \cite{Schroder2007}, 20 by \cite{Williams2014}, 22 by \cite{Said2016} and 19 by \cite{Schroder2019a}. In the available catalogue, we included the positions of the galaxies together with morphological and NIR photometric parameters identifying the mid-infrared sources of the Wide-field Infrared Survey Explorer mission (WISE; \citealt{Wright2010}). Panel (a) of Fig.~\ref{fig:all_1} 
shows the distribution of galaxies from the VVV NIRGC with the superposition of the A$_V$ contours at 10, 15, 20 and 25 mag derived from the extinction maps of \citet{Schlafly2011}.  

There are 9 sources in this region observed by Gaia-DR3 \citep{Brown2021}. 
These sources have satisfied our photometric criteria but they are faint objects (median  $K_{s}$ = 15.06 mag) near the completeness limits of the VVV NIRGC. Looking at the RGB images they look like multiple stars easily confused with extragalactic sources and they represent 0.16\% of the total number of galaxies 
in the VVV NIRGC.  In the following analysis presented in this work, we considered 5554 galaxies from the VVV NIRGC.

There are also some X-ray sources observed by the X-ray Multiple Mirror Mission 
(XMM-Newton, \citealt{Brandt2005}).
\cite{Zhang2021} applied a series of methods based on machine learning to classify these X-ray sources. The authors use panchromatic information as a training set, including data from the Sloan Digital Sky Survey (SDSS-DR12, \citealt{Eisenstein2011}), the ALLWISE database \citep{Wright2010} and the Large Sky Area Multi-Object Fiber Spectroscopic Telescope (LAMOST, \citealt{Dong2018}). Within  1 arcsec of their reported X-ray positions, we found only three VVV NIRGC galaxies. 

In the region of the VVV NIRGC, there are only eight objects with available radial velocities.
We observed the galaxy VVV-J144321.06-611753.9  
using FLAMINGOS-2 at the Gemini-South telescope \citep{Baravalle2019} measuring its radial velocity and identifying it as part of the concentration of galaxies VVV-J144321-611754.  
There are also two objects with radial velocities:  VVV-J134649.02-602429.3 and VVV-J141552.18-585535.1, previously determined by \cite{West1989} and \cite{Said2016}, respectively. The differences in the coordinates between VVV and those authors are smaller than 1 arcsec.   
Five galaxies are also 2MASX galaxies and the median angular distance with our positions is 1.725 arcmin with the highest value of 5.060 arcmin (see table B1 in \citealt{Baravalle2021} for details). 

There are 124 galaxies reported by the Parkes HI Zone of Avoidance Survey \citep{Staveley2016} in this region.  
The full width at half-power beam size of the survey is about 15.5 arcmin and it was difficult to find counterparts in this high extinction area.  
Using an angular separation of 14 arcmin, we found 95 sources of the HI survey and these high differences among sources do not allow us to associate them with VVV NIRGC galaxies.  

\section{Underlying structure behind the Southern Galactic disc}

In order to analyse the underlying galaxy structure beyond the Southern Galactic disc using the VVV NIRGC, we applied four methods.  We used the available photometry in the catalogue and the small number of objects with available spectroscopy. 
The studied methods to determine the underlying clustering 
structure are the Voronoi tessellations, Minimum Spanning Tree, the Ordering Points to Identify the Clustering, and an algorithm to identify compact groups of galaxies.  In general, they have been used in regions with moderate to low extinctions. In the regions of the VVV NIRGC, the interestellar extinction will most certainly have an effect on the analysis and estimation of the clustering properties.  

\subsection {The Voronoi tessellations}

The Voronoi tessellation \citep{Voronoi1908} is a process which can be observed often in nature. Voronoi cells can be imagined as a foam where each bubble is formed according to the positions of a given distribution of points. 
If each bubble grows at exactly the same rate during the same period, the surface area of each bubble will become limited by its neighbours, forming contact or boundary planes.
These bisecting planes will be perpendicular to the lines joining the initial centre of the bubbles/cells, and will intersect each other forming a network \citep{Icke1987}.
Thus, the edges of each cell or \emph{Voronoi polytope} are defined in such a way that any position in the area or volume of each cell is always closer to the point at the centre of the cell. 
Since the Voronoi polytopes do not assume in advance any particular binning or geometry, the detection procedure derived from these cells is highly versatile and unbiased.    
In simple words, this technique allows the detection of overdensities by analyzing the distribution of the area of the Voronoi cells. The applications of the Voronoi tessellations are varied and in particular in astronomy have been customarily applied to the analysis of the spatial distribution of galaxies in different contexts \citep[e.g.][]{Ebeling1993,Ramella2001,Sochting2002,Sochting2004,Springel2010,Lang2015}. In our analysis, we have followed the \cite{Ramella2001} prescription with the critical parameters being the inverse cell area $\tilde{f}=f/<f>$, where $f=A^{-1}$ is the inverse of the Voronoi cell area in 2 dimensions $A$, and $N$, the minimum number of galaxies that an overdensity candidate needs to have. In particular, for our analysis we have assumed a $\tilde{f}=1.2$ and $N=5$ because
they are in agreement with the mock catalogues (see the Appendix~\ref{app:voro} for details).  
 
From the VVV NIRGC in the Southern Galactic disc, we found 57 overdensities using the Voronoi tessellations. Panel (b) of Fig.~\ref{fig:all_1} 
shows the distribution of structures using the Voronoi tessellation superimposed with the A$_{V}$ iso-contours.  The overdensities (circles) have only considered at least $N=5$ adjacent cells with a $\tilde{f} = 1.2$ which have been coloured according to the area of the respective polytope (darker shades of blue for smaller areas and brown for the brightest ones). 
 
In the Appendix, Table~\ref{tab:tab_voro} shows the first  systems identified using the Voronoi tessellations
and Table~\ref{tab:tab_galvoro} presents the first galaxy members of the systems.  The systems VOR \#22 and VOR \#41 contain one galaxy with available radial velocity.

\subsection{The Minimum Spanning Tree}

The Minimum Spanning Tree (MST) has been customarily a technique of
detection and identification of overdensities for a variety of
astronomical objects \citep[e.g.][]{Campana2008}. 
The MST is a construct of graph theory which
connects all the points (objects) with an arbitrary set of positions.  
From all the possible combinations which would connect all the points in the distribution, the MST is an unique case
that minimizes the sum of the straight lines \emph{"edges''}
connecting all the points \emph{"vertices''}, and without closed
loops. 
Several algorithms have been devised which allow to obtain such
a tree, where  Kruskal's \citep{Kruskal1956} and Prim 
\citep{Prim1957} algorithms are some of the most commonly used. 
\citeauthor{Kruskal1956} algorithm takes advantage of the uniqueness of the MST by sorting all the possible edges according to their
length. Thus,  due to this characteristic, the MST must  contains
all the shortest edges which connect all the vertices to the
tree. 
\citeauthor{Prim1957} algorithm, on the other hand, starts from an arbitrary
vertex to increase the number of vertices successively one by one,
choosing at each step the shortest edge, as long as the latter does
not produce a closed loop in the tree.  
The final MST results will include all the edges in the distribution. 
However, we can limit the maximum length for the edges,
which will produce several sub-trees or a minimum spanning forest.
This limited branch length or ``fracture length'' or $D_{\textit{break}}$ of the MST traces
local overdensities which can be identified as galaxy system candidates.  
Thus, our detection will be dependent on two
parameters, the minimum number of vertices based on which we will consider an overdensity candidate, and the normalized fracture length $\tilde{D}_{\textit{break}}=D_{\textit{break}}/<D>$.

Using the MST technique with $\tilde{D}_{\textit{break}}=0.8$ in the VVV NIRGC in the Southern Galactic disc, we found 56 overdensities with at least 5 galaxies.  Panel (c) of Fig.~\ref{fig:all_1}
shows the angular distribution of structures using MST. 
In the Appendix, Table~\ref{tab:tab_MST} shows the first  systems identified using the MST technique 
and Table~\ref{tab:tab_galMST} presents the first galaxy members of the systems. The system MST \#41 contains one galaxy with available radial velocity.

\subsection{Ordering Points To Identify the Clustering Structure}

In the last years, a variety of clustering algorithms have been developed in the field of unsupervised classification methods. One of them is the Ordering Points To Identify the Clustering Structure (OPTICS) algorithm \citep{Optics}, which is a density-based method that uses a metric of distance to isolate high density regions from the surrounding environment. This algorithm has been recently used to extract clustering information in both observational \citep{McConnachie2018,Canovas2019,Massaro2019} and simulated data \citep{Fuentes2017}.

\begin{figure}
\begin{center}
\includegraphics[width=0.85\columnwidth]{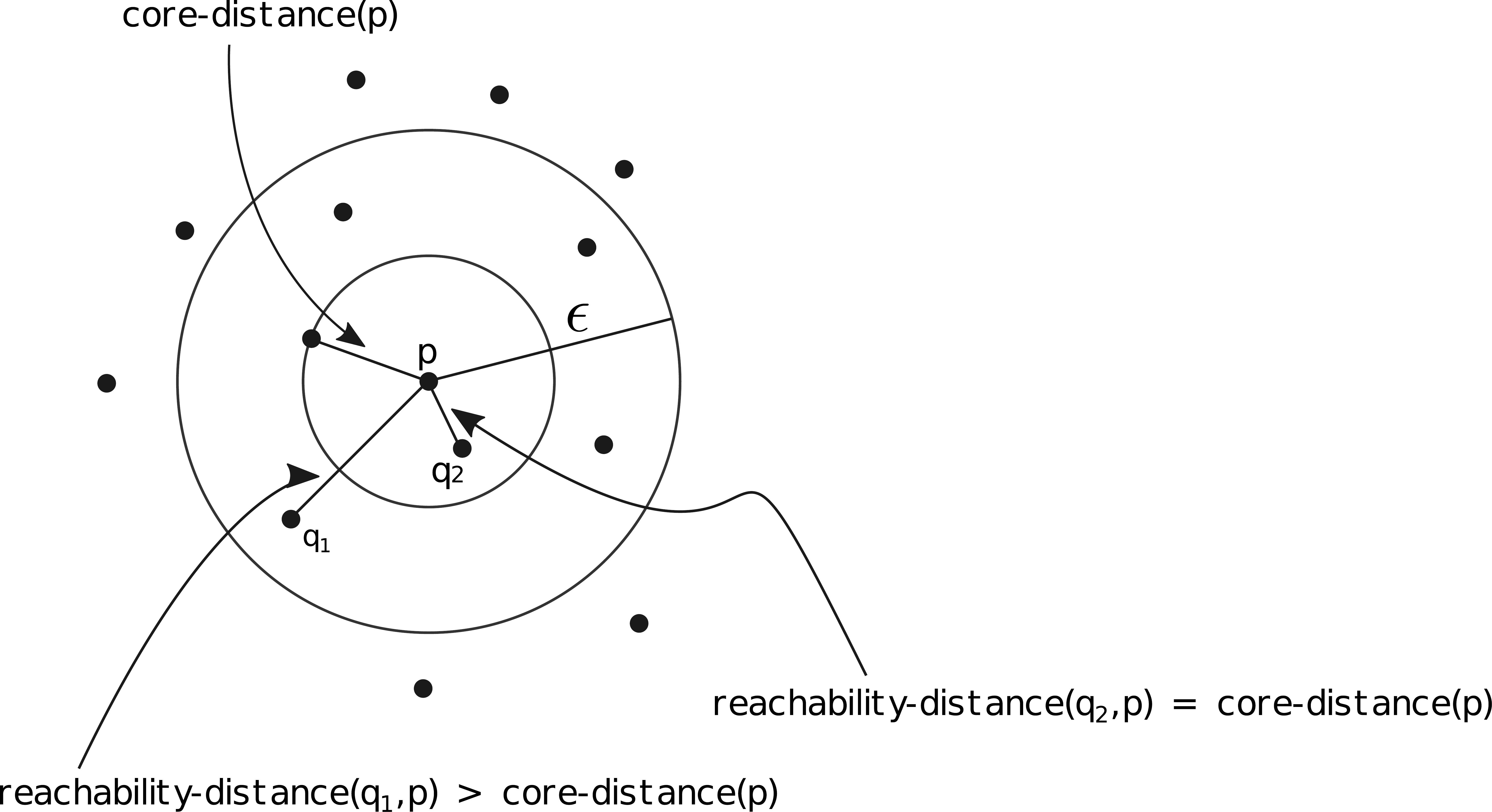}
\includegraphics[width=0.7\columnwidth]{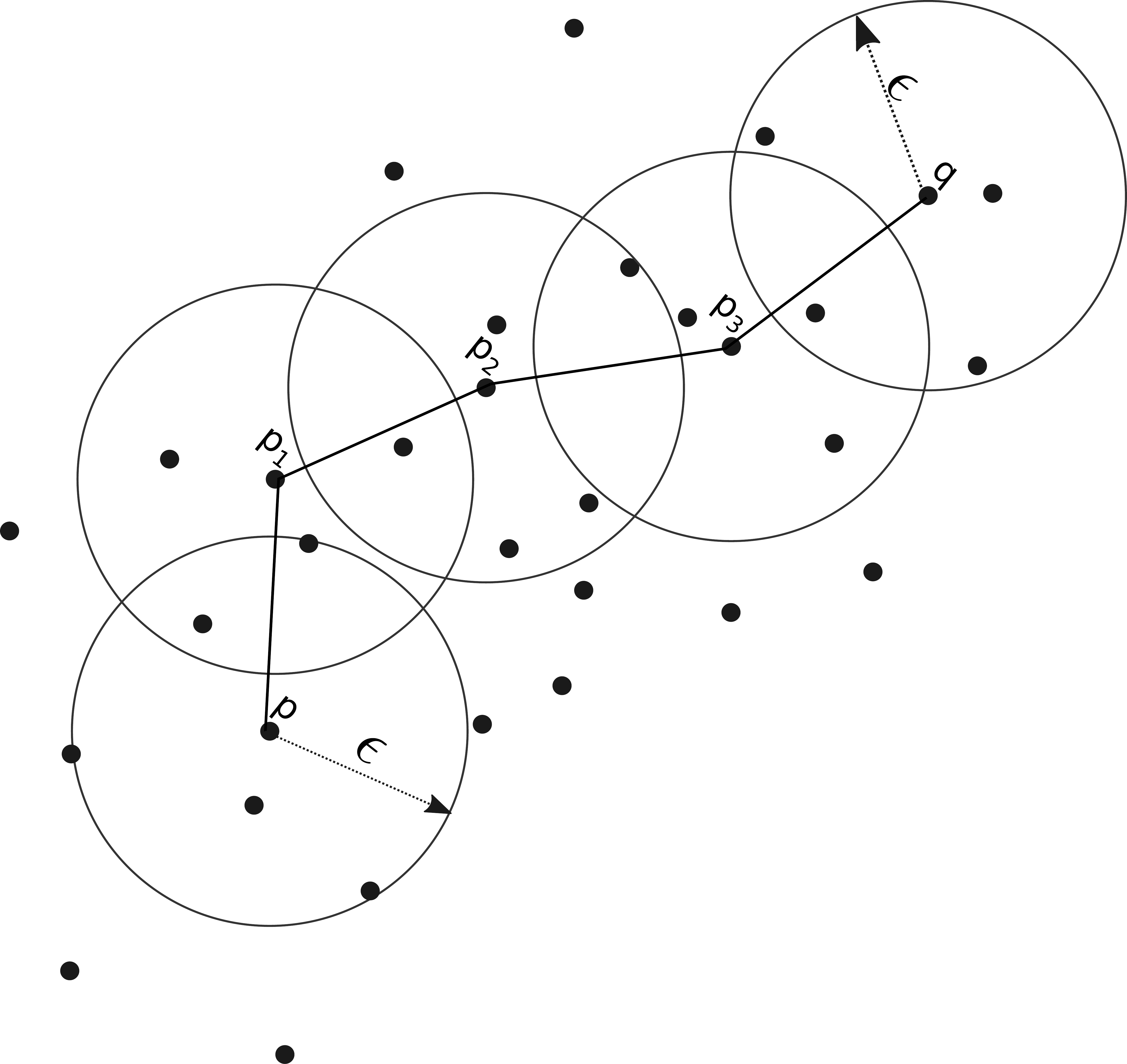}
\caption{{\bf Upper panel:} Sketch of the principal definitions used in the OPTICS algorithm. {\bf Bottom panel:} Illustration of the appended procedure in OPTICS to form clusters with the data. }
\label{fig:optics}
\end{center}
\end{figure}

The OPTICS algorithm has two main hyper-parameters to be defined: the minimum number of points $N$ 
to constitute a core-point and the maximum distance $\epsilon$ to search for neighbouring points. 
In this context, we might say that a point $p$ is a core-point if at least $N$ points (including $p$) are found within a region of radius $r \le \epsilon$ around that point. The core-distance and the {\it reachability-distance} are also important for this algorithm. The first one is defined as the smallest radius of the region which contains the $N$ points. The {\it reachability-distance} of a given point $q$ relative to the point $p$ is the maximum value between the core-distance of $p$ and the distance between $p$ and $q$.   
The maximum value that the reachability-distance could take is $\epsilon$. 
The upper panel of Fig.~\ref{fig:optics} shows a sketch representation of these concepts for the case of $N = 3$. In this plot, $p$ is considered the core-point and the core-distance is given by the radius enclosing $N$ points. 
Following the previous definitions, the reachability-distance of the points $q_1$ and $q_2$ is the distance relative to the point $p$ and the core-distance of $p$, respectively. 

The identification procedure starts taking an arbitrary point $p$ and looking for the neighbouring points within a region of radius $\epsilon$. If $p$ is a core-point (as defined above), then all density-reachable points from $p$ are appended to this and a cluster\footnote{We will use the word cluster as it is commonly used in the clustering algorithm language, and which can be understood, in our case, as an over-density in the projected distribution of galaxies.} is formed.  
The bottom panel of Fig.~\ref{fig:optics} illustrates how a cluster is built through the process of appending density-reachable points. For the sake of clarity, we might also say that a point $q$ is a density-reachable point relative to a point $p$ if there is a path of points $p_1$, $p_2$, ..., $p_n$ such that $p_{i+1}$ lies within a region of radius $r \le \epsilon$ centred on $p_i$. The particularity that makes OPTICS different from other density-based methods is that it uses both the core-distance and the reachability-distance to order the point sample and to construct the reachability-plot which characterizes the density structure of the data.  

By construction, OPTICS can be applied to any n-dimensional data set for which a metric of distance can be defined. In our case, and since we are only using angular spatial data, we choose the angular separation as the distance. For our purposes, we fix $N = 4$ while a value of $\epsilon = 0.07$ is chosen by eye inspection of the reachability-plot in order to only keep the denser systems minimizing the background noise. These adopted values are in agreement with the results using mock catalogues (see the Appendix \ref{app:optics} for details). 

Applying  this method to the VVV NIRGC galaxies in the Southern Galactic disc,  we identified 98 overdensities with 5 or more members. 
Panel (d) of Fig.~\ref{fig:all_1}
shows the angular distribution of these galaxy systems using OPTICS. In the Appendix, Table~\ref{tab:tab_optics}  shows the first  systems identified using the OPTICS algorithm 
and Table~\ref{tab:tab_galoptics} presents the first galaxy members of the systems. The systems OPT \#51 and OPT \#68 contain one galaxy with available radial velocity.

\subsection {Compact groups}

We also used the VVV NIRGC to identify compact groups (CGs) as defined by \cite{Hickson82}, and using a slightly different procedure than that used by \cite{DiazGimenez+12} in the 2MASS catalogue.
Briefly, the algorithm searches for galaxy associations in projection on the sky that fulfil all the following criteria:

\begin{description}
\item[\bf Membership:] the number of galaxies with apparent magnitudes brighter than $K_{s_{\rm b}} + 3$ (where $K_{s_{\rm b}}$ is the extinction-corrected apparent magnitude of the brightest galaxy in the group) must not be below a minimum or exceed a maximum (see below for these limits).
The magnitude range of three magnitudes from the brightest galaxy was originally thought to avoid the presence of many interlopers in projection;

\item[\bf Isolation:]  there are no other galaxies 
with magnitudes brighter than $K_{s_{\rm b}} + 3$ within a disk with radius equal to three times the angular radius of the minimum circle that encloses all the galaxy members;

\item[\bf Compactness:] the mean surface brightness of the group, computed by adding the flux of all the galaxy members averaged over the minimum circle, 
is brighter than a given threshold;

\item[\bf Flux limit:] the brightest member has to be three magnitudes brighter than the limit of the catalogue to ensure that the membership and isolation account for all possible neighbours in the range of three magnitudes from the brightest one.
\end{description}

The original \cite{Hickson82}'s sample had CGs with four to ten members. However, when radial velocities became available, concordant triplets were allowed in the sample \citep{Hickson92}. Also, \cite{McConnachie+09} and \cite{Zheng+20} have introduced CG catalogues 
including triplets. 
Following these works and allowing to have triplets among the CGs, we adopted a minimum number of three galaxies and a maximum of ten. 
Notice that membership and isolation are computed only with galaxies within a range of three magnitudes from the brightest one. Fainter galaxies lying in the same region are not considered as galaxy members and therefore do not violate the isolation criteria.  The surface brightness threshold adopted in this work in the $K_{s}$ passband is the same adopted for the 2MASS CG sample, i.e., $\mu_{Ks}\le 23.6 \, \rm mag/arcsec^2$. 

\begin{figure}
    \centering
    \includegraphics[width=0.48\textwidth]{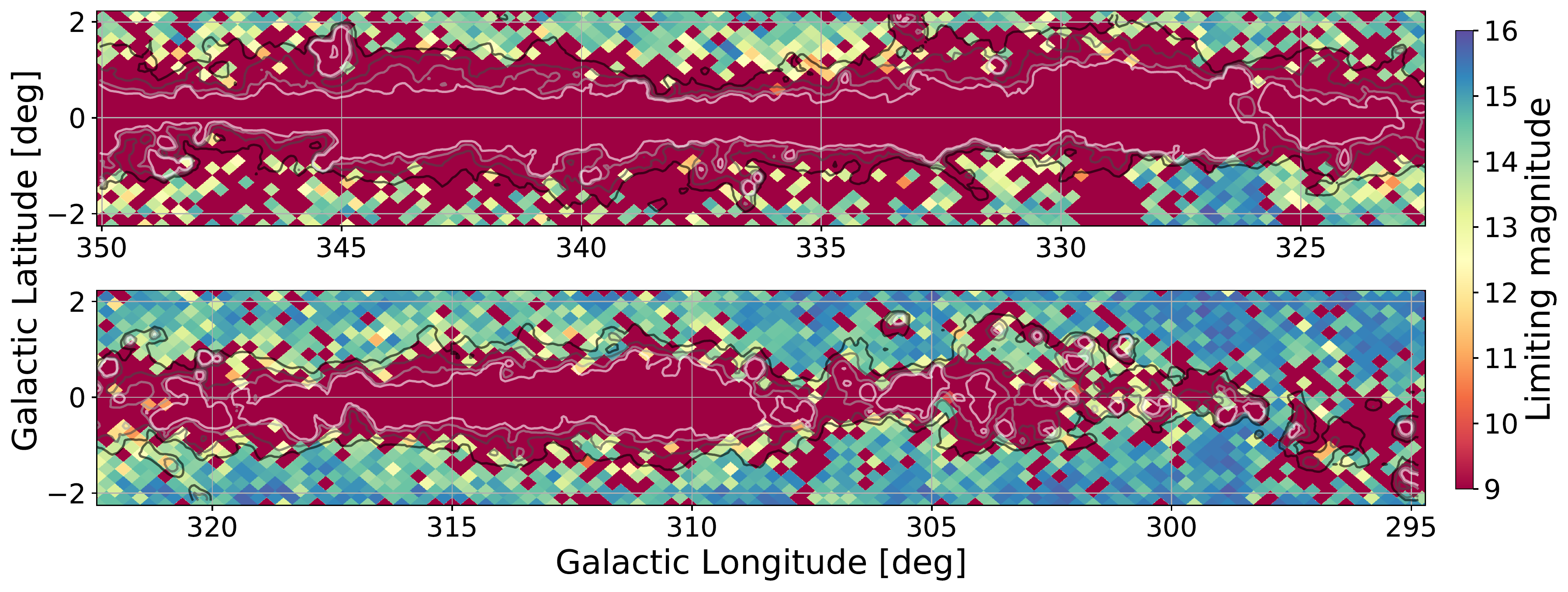}
    \caption{$K_{s}$ magnitude limit mask in the regions of the VVV NIRGC.  The colours are related to the limiting magnitude and only galaxies brighter than the associated limit can be observed in each pixel.
    }
    \label{fig:mask_mag}
\end{figure}
Regarding the flux limit of the brightest galaxy, the original algorithm is thought to be applied to a complete flux-limited sample of galaxies. When examining the distribution of extinction-corrected $K_{s}$ magnitudes of the VVV NIRGC galaxies, a decrease in the number of galaxies towards the limit of the catalogue is observed (fig.~6 in \citealt{Baravalle2021}). The VVV NIRGC is nearly complete up to a magnitude of 14-14.50 mag, and then it starts departing from the expectation, i.e., there is no uniform extinction-corrected apparent magnitude limit in the studied area of the sky. 

Figure~\ref{fig:mask_mag} shows the angular magnitude limit mask according to the $K_{s}$ apparent magnitude of galaxies in the VVV NIRGC. 
The radius of each pixel in this mask is approximately $7 \, \rm arcmin$. Each pixel is associated with a magnitude  ($Ks_{\rm max}$) corresponding to the faintest galaxy that was observed in this pixel. The lack of uniformity is due partially to the intrinsic interstellar extinction and  to the stellar density in the Zone of Avoidance \citep{Baravalle2021}, but also to inhomogeneities caused by the visual inspection of the images.
Therefore, the survey magnitude limit varies with position on the sky, and this should be considered when identifying the CGs. 
We implemented two different approaches. 
First, taking into account the pixel position of the brightest galaxy, the 
flux limit criterion was applied as $Ks_{\rm brightest} \le (Ks_{\rm max} - 3)$.
This is a {\it restricted} search in which we can assure that all the CGs accomplish all the four criteria mentioned above. 
Second, we performed a more {\it unrestricted} search in which we do not apply the flux limit criterion at all. This method was actually followed by \citeauthor{Hickson82} himself, and also applied to the SDSS galaxies by \cite{McConnachie+09}. Not imposing the limit on the brightest galaxy makes membership and isolation questionable (see \citealt{DiazGimenez+10} and \citealt{McConnachie+09} for further details). However, they might still be overdensities on the sky.

The basics of this implemented algorithm are shown in fig.~1 of \cite{DiazGimenez+18} (the first block of the left flow chart).
The {\it restricted} search leads us to a sample of $4$ CGs in the VVV NIRGC, while the {\it unrestricted} search produces a sample of $81$ CGs (that includes the other $4$). Among these groups, we find $70$ triplets, $9$ quartets, $1$ quintet, and $1$ sextet. 
Panel (e) of Fig.~\ref{fig:all_1} 
shows the angular distribution of the {\it unrestricted} CGs. 
In the Appendix, Table~\ref{tab:tab_cg} shows the CGs identified in the VVV NIRGC. The first four rows correspond to the 4 CGs identified in the {\it restricted} search for which we can assure that membership and isolation are appropriately verified, while the remaining identified with the {\it unrestricted} algorithm are ordered by Galactic latitude. Table~\ref{tab:tab_galcg} presents the galaxy members of these systems. The CG \#48 contains one galaxy with available radial velocity.

\section{Main results}

Using the different clustering methods, we obtained 57 systems with Voronoi tessellation, 56 with MST technique, 98 with OPTICS considering 5 or more galaxies, and 81 compact groups with 3 or more members.  Four system samples were generated and these systems are shown in the Tables presented in Appendix~\ref{galaxysystems}. Only eight of these systems have  one galaxy with available redshift information in the literature.  

\begin{figure}
\begin{center}
\includegraphics[width=0.45\textwidth]{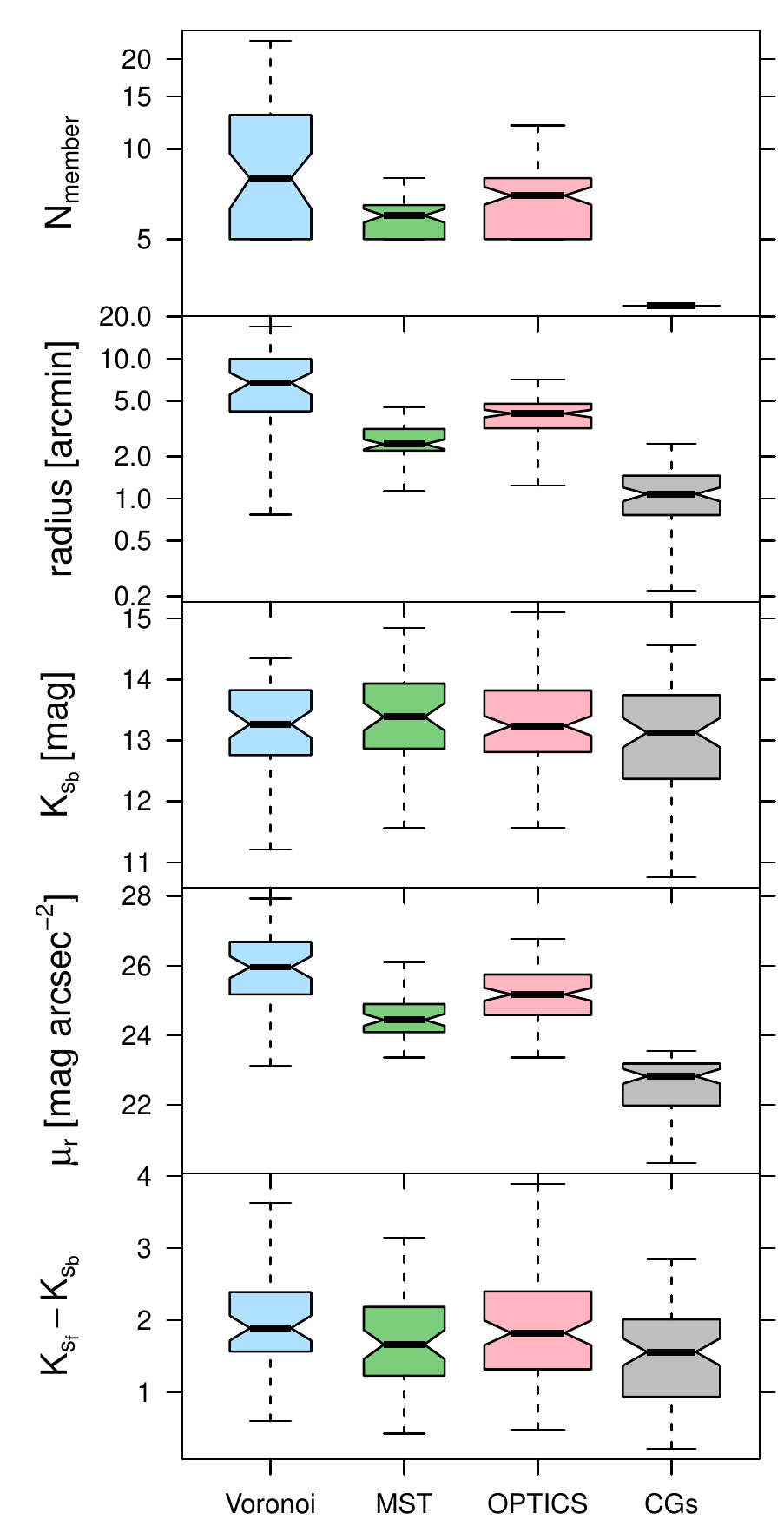}
\caption{Distributions of the main system properties as boxplots.  The properties, from top to bottom, are number of galaxy members, radius of the minimum circle that encloses the galaxy members, extinction-corrected apparent magnitude of the brightest galaxy of the group, mean surface brightness of the group, and difference of apparent magnitudes between the faintest and the brightest galaxy members. The \emph{waists}, tops and bottoms of the \emph{coloured boxes} indicate the median values, and the 75th and 25th percentiles, respectively. The \emph{widths} of the boxes are proportional to the numbers of systems in each sample, while \emph{notches} around the waists show the $95\%$ confidence interval on the medians. }
\label{fig:comparando2}
\end{center}
\end{figure}

Figure~\ref{fig:comparando2} shows the main properties of these four system samples, which are the  
number of galaxy members, the radius of the minimum circle that encloses the galaxy members, the extinction-corrected apparent magnitude of the brightest galaxy, the mean surface brightness of the system averaged over the minimum circle, and the magnitude range of the galaxy members.
Some of the observed differences in these properties are directly related to the algorithm themselves. The minimum number of members is one of the inputs in each algorithm, ranging from 3 members in a given range of magnitudes for CGs, to 5 galaxies for Voronoi, MST and OPTICS systems. The maximum number of members is not an input parameter.
Voronoi systems have the highest number of members followed by MST and OPTICS. 
The CGs are obviously the least numerous systems.  
Also, by definition, CGs groups depart from the rest in their smaller angular radius and greater compactness (lower values of $\mu_k$), followed by the MST, then OPTICS and finally Voronoi systems.
There are no significant differences in the magnitudes of the brightest galaxies nor the magnitude ranges of the galaxy members (the notches of the boxplots overlap).

\begin{table}
\caption{Percentage of systems obtained with the four methods (rows) that share galaxies with systems in other method (columns).}
    \label{tab:matching}
    \centering
    \begin{tabular}{rccccc}
    \hline
               & Voronoi  & MST       & OPTICS    & CGs     & No-match \\
    \hline
    Voronoi    & --       &   65\%    &  81\%   &   42\%  & 14\%  \\
    MST        &  84\%    &   --      &  98\%    &   36\%  & \ 2\% \\
    OPTICS     &  64\%    &   55\%    &   --      &   31\%  & 21\%\\
    CGs        &  38\%    &  26\%     &  42\%   &    --   & 49\%  \\
    \hline
    \end{tabular}
\end{table}

\begin{figure*}
   \begin{minipage}{0.5\textwidth}
     \centering
    \includegraphics[width=1\linewidth]{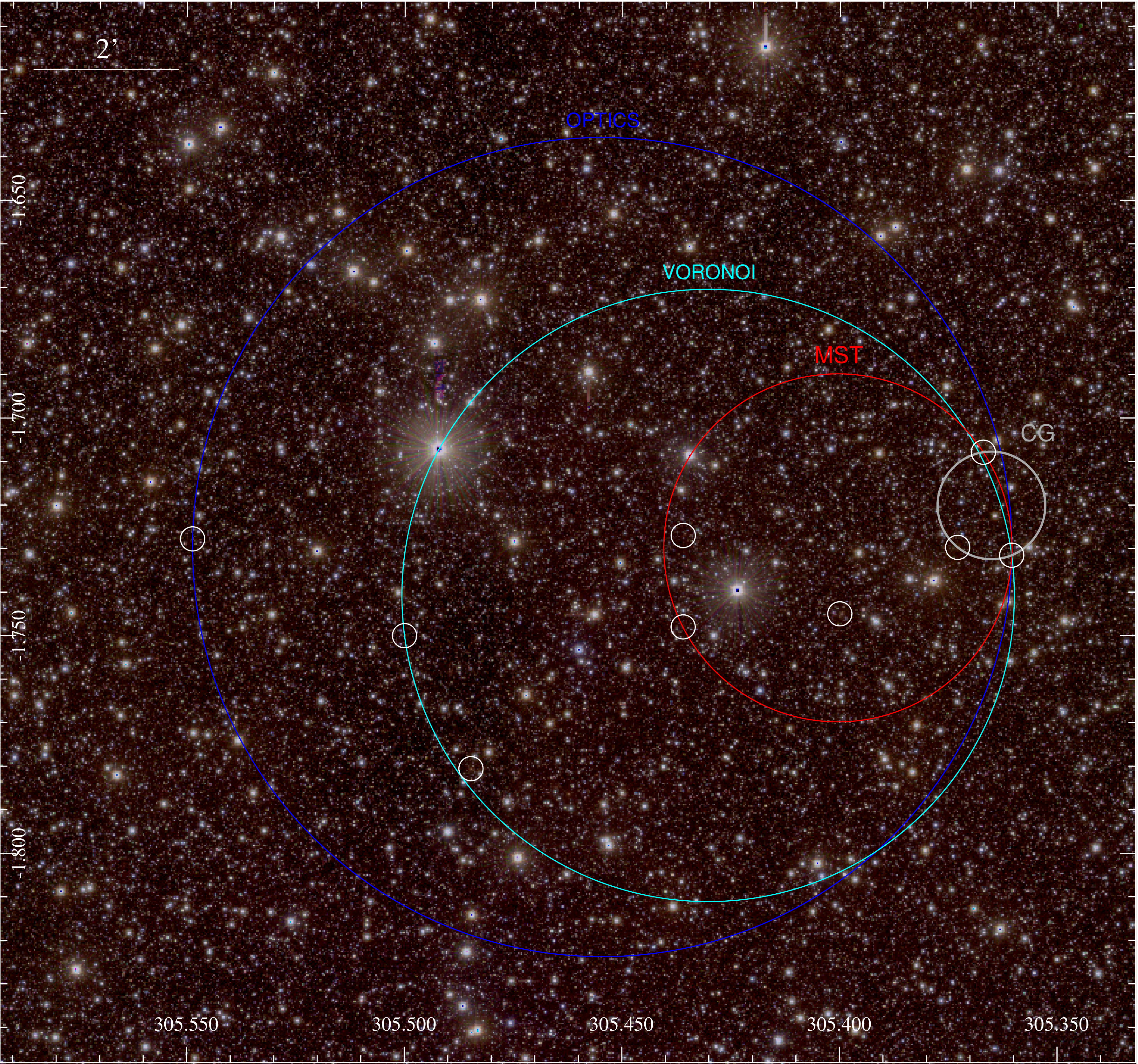}
   \end{minipage}\hfill
   \begin{minipage}{0.5\textwidth}
     \centering
     \includegraphics[width=1.0\linewidth]{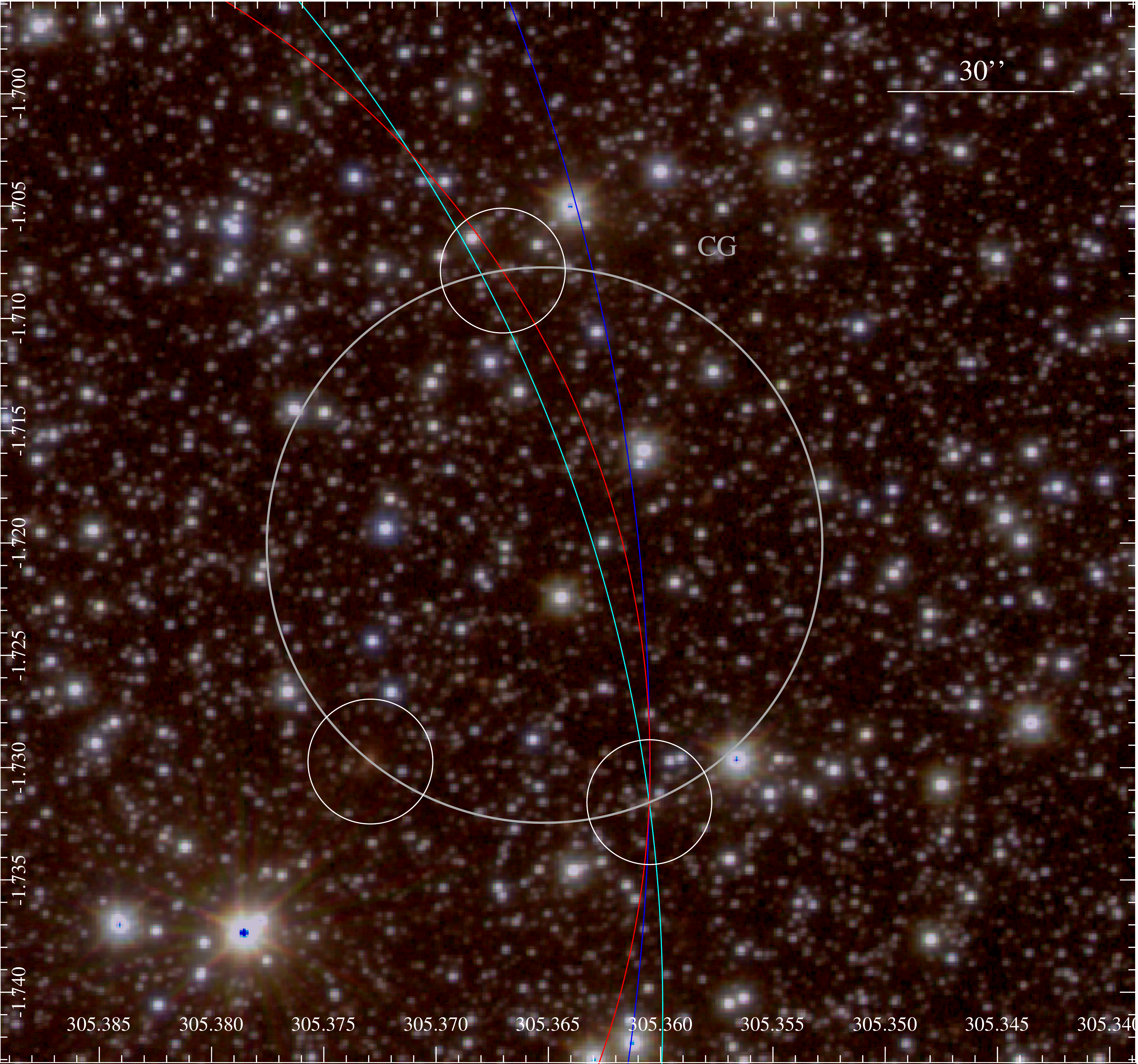}
   \end{minipage}
\caption{Galaxy system example (VOR \#29, MST \#22, OPT \#36 and CG \#27) 
identified by the four methods.  Colour images showing the region of 16 $\times$ 16 arcmin$^{2}$ (left panel) and a zoom around the CG of 3 $\times$ 3 arcmin$^{2}$ (right panel). Solid circles represent the minimum circle that encloses the galaxy members defined by the methods.
}
\label{fig:4match}
\end{figure*}

In addition, some of the systems identified with the different methods belong to more than one sample. We perform a member-to-member comparison, and we consider a ``match'' when systems share at least one galaxy member. This definition is not restrictive due to the different nature of the used algorithms (for instance, the magnitude range used by one algorithm differs from the others). Therefore, this might be considered as a lower limit to the group matching rate. 
Table~\ref{tab:matching} quotes, in each row, the percentage of a given sample that matches with systems in the other samples or without matching (last column).  We found nineteen systems that are identified with the four methods at the same time and share at least two members (68\% share three or more members). 
These overdensities have higher probabilities to be real associations. 
The galaxy cluster VVV-J144321-611754 \citep{Baravalle2019} is one of these systems, which is identified as VOR \#41, MST \#41, OPT \#68 and CG \#48. Fig.~\ref{fig:4match} shows an example (VOR \#29, MST \#22, OPT \#36 and CG \#27) of these quadruple identifications. 
The left panel is a 16 $\times$ 16 arcmin$^2$ composite ($J$, $H$, and $K_{s}$) colour image of the field. The circles represent the minimum circle that encloses the galaxy members (solid coloured circles)  obtained by the four methods.  
The right panel is a zoom of 3 $\times$ 3 arcmin$^2$ around the CG centre.  Also in Appendix~\ref{galaxysystems}, Fig.~\ref{fig:cuadruple} shows the nineteen overdensities identified by the four algorithms in projection. 
It can be seen that the different algorithms used in this work are looking for associations that enclose different overdensities. The CGs and, in particular, the unrestricted CGs might be high overdensities that are actually part of larger structures.  \cite{Taverna2022}  showed recently that around 90\% of the Hickson-like CGs in redshift-space are not isolated.

\subsection{Assessing the probability of finding true associations}

Since the overdensities presented in this work are identified in projection, one may wonder the real probability that these systems are formed by concordant galaxies (in redshift space). To answer this, we rely on the results obtained from mock catalogues (see Appendix~\ref{app:mock} for details).  We identified overdensities in these mock catalogues in the same way as in the VVV NIRGC and we applied a velocity filtering in redshift space. 

We explored the best input parameters in the four different clustering methods 
 to be used in the VVV NIRGC.  A median\footnote{We compute the median over the 30 mock catalogues. We assign as error the 95\% confidence interval on the medians.} value of 
$(30 \pm 6)\%$ for the Voronoi overdensities remain with more than $5$ concordant members and a $\tilde{f}=1.2$.  For the MST overdensities, $(28 \pm 5)\%$ remain with more than $5 $ concordant members and a $\tilde{D}_{\textit{break}}=0.8$.  For OPTICS, 
$(20 \pm 3)\%$ remain with more than $5$ concordant members.
Finally, $(46 \pm 5)\%$ of the CGs identified with the {\it restricted} algorithm survive the velocity filtering, while this percentage is $(33 \pm 1)\%$ for CGs identified with the {\it unrestricted} algorithm. 



These predictions translated to the overdensities identified in the VVV NIRGC would indicate that between $14$ and $21$ of the Voronoi results would survive the velocity filtering.  For the cases of MST and OPTICS, between $13$ and $18$, and $16$ and $23$ of the overdensities, respectively.  Only $1$ or $2$ of the {\it restricted} VVV CGs could be true associations in redshift-space, while approximately $27$ of the {\it unrestricted} VVV CGs would have at least three concordant members. Assuming that the mock catalogs (described in Appendix~\ref{app:mock}) are representative of the VVV NIRGC, we can predict that the redshifts of the overdensities might be in the range [0.005,0.450], with a median redshift of the around $z\sim 0.1$. 

In all cases, the fraction of possible chance alignments in the VVV NIRGC is significant, which  needs
to be confirmed by follow-up redshift measurements. Due to the high stellar extinction in this region, a suitable instrument for performing spectroscopic observations could be one like the near-infrared imaging spectrograph, such as Gemini-South,  Flamingos-2.



\section{Summary}

The VVV Near-IR Galaxy Catalogue contains 5563 galaxies beyond the Southern Galactic disk. We applied four clustering methods to detect overdensities with the 5554 galaxies in the catalogue, namely the Voronoi tessellations; the Minimum Spanning Tree and the Ordering
Points To Identify the Clustering Structure; and compact groups of galaxies.
In projection, we obtained 57 systems with Voronoi tessellation, 56 with MST technique, 98 with OPTICS considering 5 or more galaxies, and 81 compact groups with 3 or more members.  We found that Voronoi systems are the biggest with the highest number of members.  The MST and OPTICS overdensities are intermediate systems.
CGs groups have the smallest angular radii and greatest compactness. These overdensities are considered potential systems of galaxies and 
nineteen of them have higher probabilities because they were detected by the four algorithms.   

Mock galaxy lightcones built from the Millennium I simulation combined with a semi-analytic model of galaxy formation are used to predict the probability of the systems identified in projection on the sky to be true associations in redshift space. We found that, depending on the algorithm, between 20 to 33 per cent of the systems identified might be populated by concordant galaxies in redshift space. 

Although we do not have enough redshift information (only eight galaxies) in the VVV NIRGC, we can infer from the mock catalogs, that the redshift distribution of the systems identified by the algorithms would be in the range of 0.005 to 0.450. 
It is fundamental for LSS studies to have spectroscopic data of the galaxies in these regions.  
Furthermore, the redshift information will allow us to disentangle members from background sources included  in projection by the different searching algorithms.

\section*{Acknowledgments}

We would like to thank the anonymous referee for the useful comments and suggestions to improve this paper. 
We gratefully acknowledge data from the ESO Public Survey program ID 179.B-2002 taken with the VISTA telescope, and products from the CASU. This research has made use of the NASA/IPAC Infrared Science Archive, which is funded by the National Aeronautics and Space Administration and operated by the California Institute of Technology.  We thank B. Henriques et al. for making their model publicly available. The Millennium Simulation databases used in this paper and the web application providing online access to them were constructed as part of the activities of the German Astrophysical Virtual Observatory (GAVO). D.M. gratefully acknowledges support by the ANID BASAL project FB210003. J. L. N. C. is grateful for the financial support received from the Southern Office of Aerospace Research and Development of the Air Force Office of the Scientific Research International Office of the United States  (SOARD/AFOSR) through grants FA9550-18-1-0018 and  FA9550-22-1-0037. 

This work was partially supported by Consejo de Investigaciones Cient\'ificas y T\'ecnicas (CONICET) and Secretar\'ia de Ciencia y T\'ecnica de la Universidad Nacional de C\'ordoba (SecyT). DM is supported by the BASAL Center for Astrophysics and Associated Technologies (CATA) through grant AFB 170002.


\section*{Data Availability}

The VVV NIRGC used in this article is available at
\url{https://catalogs.oac.uncor.edu/vvv\_nirgc/}. 
System catalogues presented in this work are available electronically. 
The derived data generated in this research will be shared on reasonable request to the corresponding authors.



\bibliographystyle{mnras}
\bibliography{Voronoi_source.bib} 


\appendix
\section{Mock galaxy catalogues}
\label{app:mock}

Following the  procedure of \cite{zeta+14}, we built an all-sky lightcone using the synthetic galaxies of the publicly available semi-analytic model of galaxy formation \citep{Henriques+20}, run on the Millennium I simulation \citep{Springel+05} and re-scaled to the Planck cosmology \citep{Planck+16}.  
Briefly, the lightcone construction takes into account the evolution of structures and galaxy properties because galaxies are extracted at different redshifts from the latest 16 outputs of the simulation.
Absolute magnitudes were interpolated between different snapshots, and are used to compute the observer-frame apparent magnitudes, including a k-decorrection procedure described in \cite{DiazGimenez+18}. 
Redshifts were computed from the comoving positions and peculiar velocities provided in the simulations. We adopted an observer frame extinction-corrected $K_{s}$ apparent magnitude limit of $15.85$ mag, which is the magnitude limit of the VVV NIRGC \citep{Baravalle2021}.

\begin{figure}
    \centering
   \includegraphics[width=0.45\textwidth]{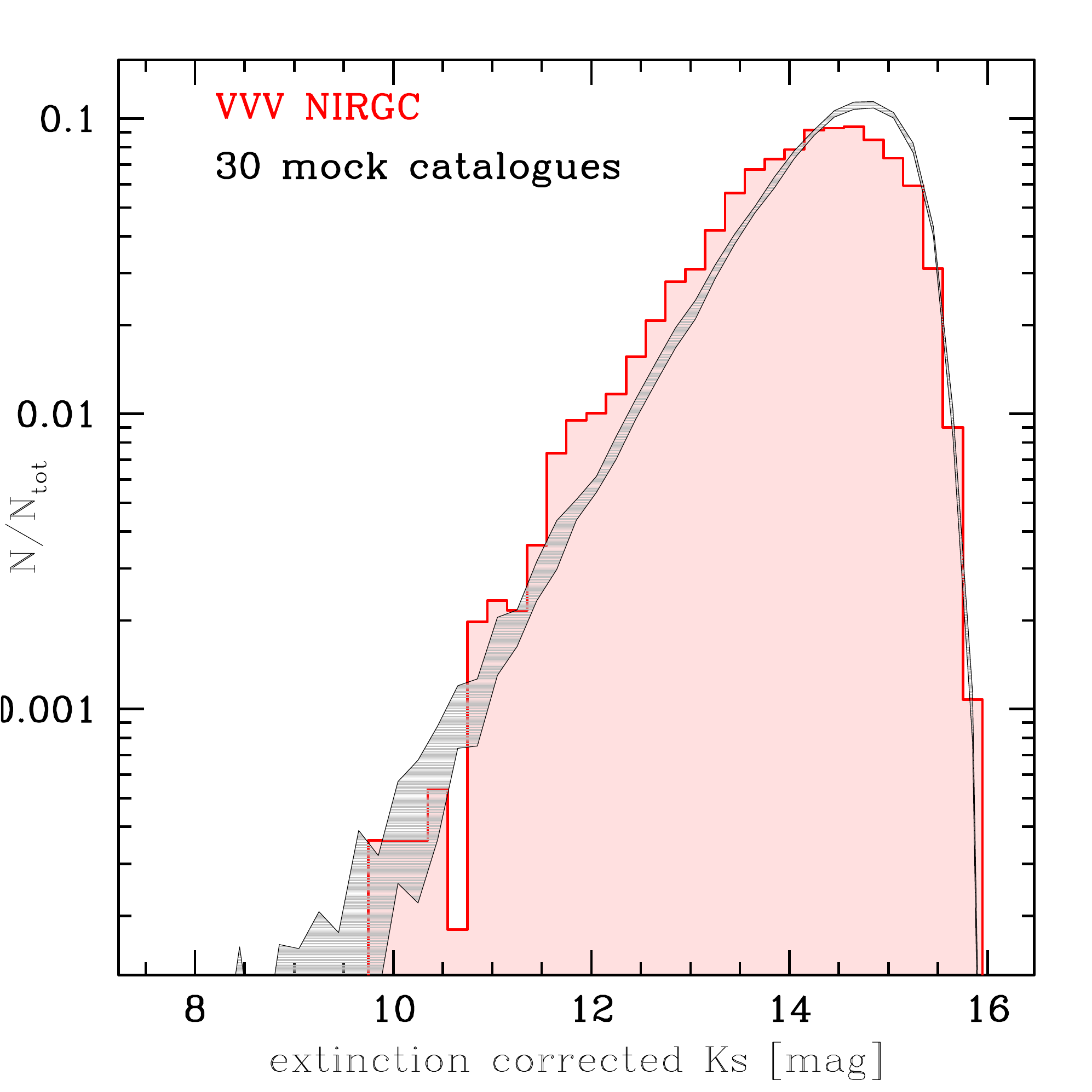}
    \caption{Normalized distributions of extinction-corrected apparent $K_{s}$ magnitudes. Red shaded histogram corresponds to galaxies in the VVV NIRGC, while the grey curve corresponds to the area delimited by the 25-th and 75-th percentile of the distribution of galaxies in 30 mock galaxy catalogues.}
    \label{fig:mag_mock}
\end{figure}
From this all-sky lightcone, we produced 30 mock galaxy catalogues pointing towards different directions in the  sky and using the angular magnitude limit mask (see Fig.~\ref{fig:mask_mag}) to reproduce the angular size and magnitude distribution of galaxies in the VVV NIRGC. 
This mask takes into account the lack of uniformity due partially to the intrinsic interstellar extinction in the ZoA, and the stellar crowding that might affect the galaxy identification in the VVV NIRGC. The number of galaxies in the mock catalogues varies from $7\,750$ to $9\,690$, i.e., a median of $1.6$ times the number of galaxies in the VVV NIRGC. This factor may be taken into account when comparing projected number densities in the mock catalogue with those in the observational catalogue. 

Figure~\ref{fig:mag_mock} shows the normalized distributions of extinction-corrected apparent magnitudes of galaxies in the VVV NIRGC and the mock catalogues. Despite the use of the magnitude mask, it can be seen that there is still an excess of faint galaxies in the mock catalogues compared to the observational data.

\subsection{Determining the best finding parameters}
\label{app:filter}

Galaxy systems found by different algorithms are dependent on the algorithms themselves and at the same time, each algorithm depends on different parameters. Mock galaxy catalogues are a useful tool for determining the set of parameters that produces the most reliable sample of galaxy systems for each finder algorithm. 
Making use of the redshift information of galaxies in the mock catalogue, it is possible to determine the probability that systems identified in projection are formed by concordant galaxies (in redshift space).  

Once the systems were identified in projection in the mock catalogues, we applied an iterative velocity filter to discard interlopers, similarly as \cite{Hickson92}.  The aim is to find the best finding parameters that maximize the survival of systems with concordant galaxies. The iterative velocity filter is as follows.  First, we define the centre of the system in the radial direction as the bi-weighted median redshift of galaxies in each system, $z_{cm}$.
Galaxies with redshifts $z_i$ satisfying
 
$$ \displaystyle  \frac{|z_i-z_{cm}|}{1+z_{cm}}> \frac{1\,000 \ \rm km/s}{c},$$
with $c$ the speed of light, are considered interlopers and discarded from the system.  We repeat the process until there are no further interlopers or the number of remaining members is less than the limit used in each method and the entire system is discarded, whichever occurs first. 
For each method and for different sets of input parameters, we computed the fraction of systems identified in projection that survive the velocity filter.

\subsubsection{Voronoi }
\label{app:voro}

\begin{figure}
\begin{center}
\includegraphics[width=0.48\textwidth]{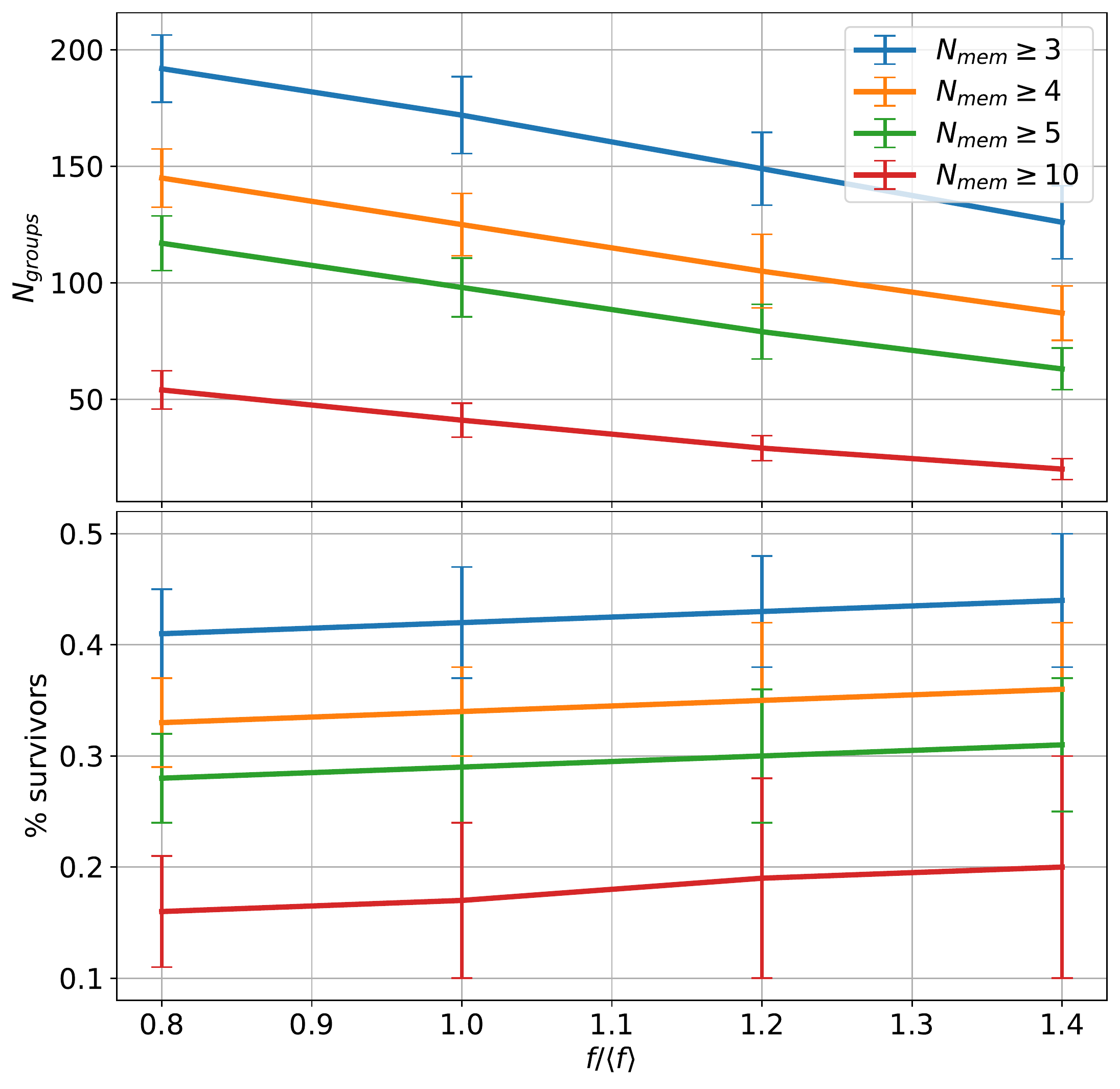}
\caption{Results of the Voronoi  algorithm in the mock catalogues. \emph{Top}, average number of system candidates selected as a function of the parameter $\tilde{f}=f/<f>$ that have at least a number $N_{mem}$ of members and in which all those have passed the iterative velocity filter.  \emph{Bottom}, ratio of overdensities that survive the velocity filter with respect to those selected for the numerical algorithm disregarding the velocity information from the mock catalogues. 
\label{fig:mock_v}}
\end{center}
\end{figure}

Figure~\ref{fig:mock_v} (upper panel) shows the results of the Voronoi  algorithm in the 30 mock catalogues as a function of the critical parameter $\tilde{f}=f/<f>$.  Each colour represents the minimum number of members $N_{mem}$ of the overdensities that have passed the iterative velocity filter, where errorbars in each case have been derived from the standard deviation of the results of all mock catalogues. Similarly, bottom panel of the figure shows the percentage of overdensities that survive the velocity filter with respect to the total number of overdensities detected by the Voronoi algorithm, not taking into account the velocity filter. Hence, assuming that the mock catalogues are a fair representation of the underlying cosmological structures behind the Galactic disk, we can estimate the number of real systems in the overdensities detected by our algorithm, without velocity information. In this sense, our results indicate that for a $\tilde{f}=1.2$ and at least $Nmem=5$ members, $30 \pm 6 \%$ of those overdensities detected by the Voronoi s in the real catalogue should be actual galaxy systems.

\subsubsection{MST algorithm}
\label{app:mst}

\begin{figure}
\begin{center}
\includegraphics[width=0.48\textwidth]{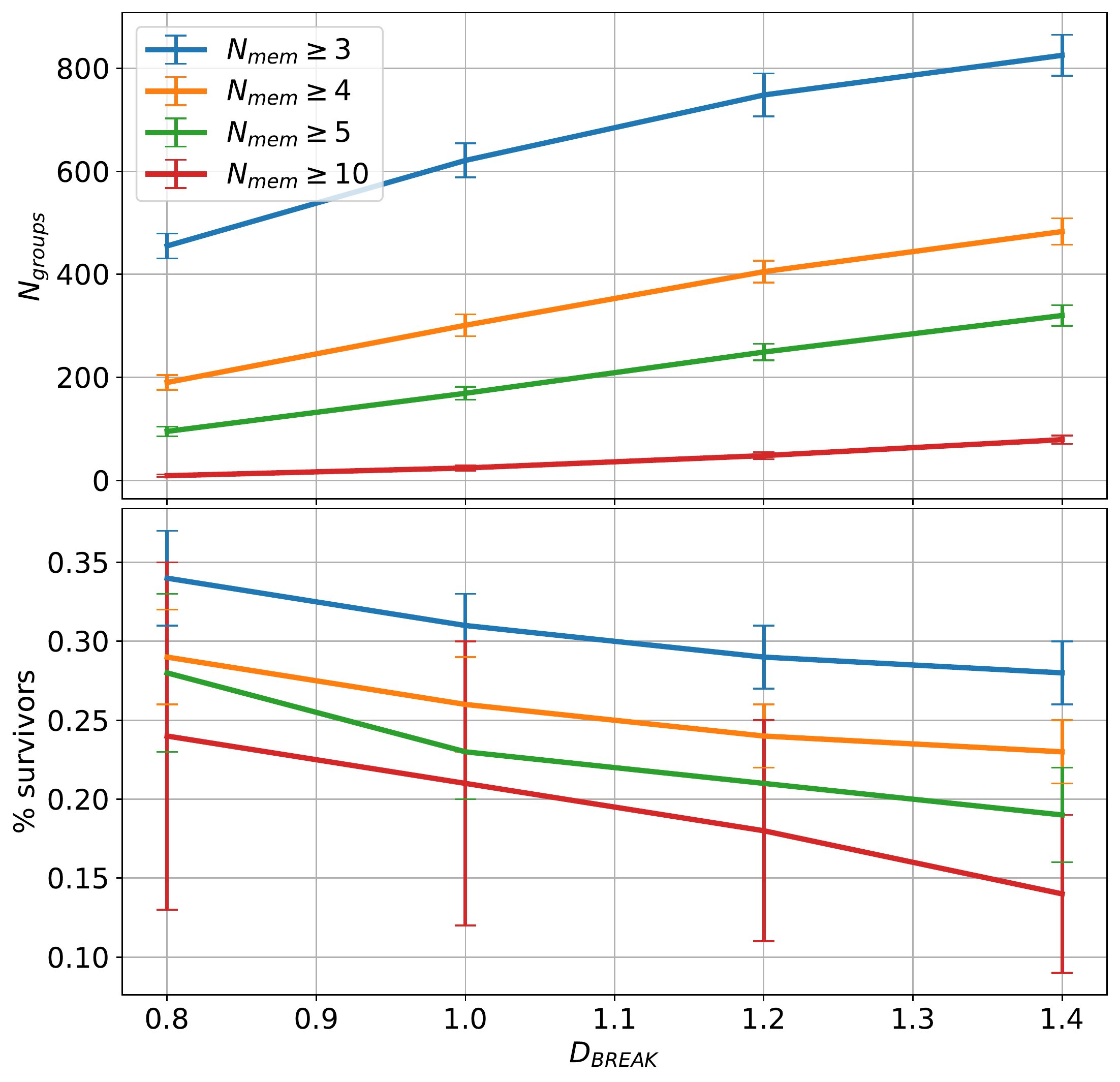}
\caption{Results of the MST algorithm in the mock catalogues.  Same as Fig.~\ref{fig:mock_v} but for the MST, the critical parameter is the normalized branch length $\tilde{D}_{\textit{break}} = D_{\textit{break}}/<D_{\textit{break}}>$. 
\label{fig:mock_mst}}
\end{center}
\end{figure}

Figure~\ref{fig:mock_mst} shows the results of the  MST algorithm in the 30 mock catalogues as a function of the normalized branch length $\tilde{D}_{\textit{break}}=D_{\textit{break}}/<D_{\textit{break}}>$,  similar to that of the Voronoi s. Just as in Fig.~\ref{fig:mock_v} the colours indicate the minimum number of members $N_{mem}$ that an overdensity should have to be considered as such. Based on these results, it is possible to estimate that for the detected overdensities with the MST algorithm using $N_{mem}=5$ and a $\tilde{D}_{\textit{break}}=0.8$, the percentage of $28 \pm 5\%$ should be actual galaxy systems.  

\subsubsection{OPTICS}
\label{app:optics}

We applied OPTICS on the 30 mock catalogues described above (Sect.~\ref{app:mock}) for a set of 11 different values of $\epsilon$ ranging from $0.02$ to $0.12$. Additionally, we explored how our results change when we decrease the minimum number of members $M_{mem}$ a cluster must contain in order to be taken into account.

\begin{figure}
\begin{center}
\includegraphics[width=0.45\textwidth]{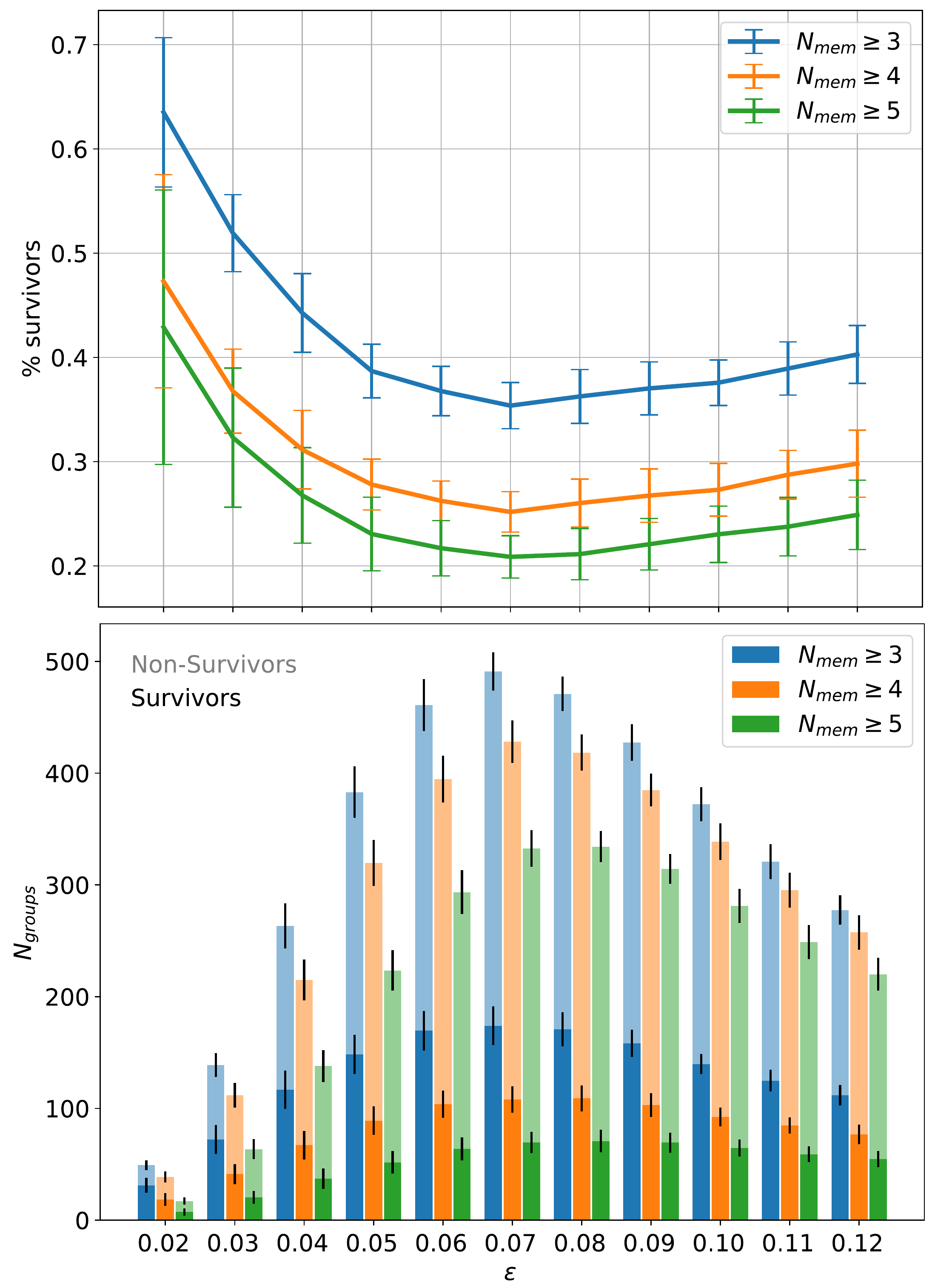}
\caption{
Upper panel shows the percentage of OPTICS surviving groups respect to the total number of identified overdensities as function of $\epsilon$ parameter. Each curve correspond to a different minimum number ($N_{mem}$) of system members.
In the bottom panel, dark bars represent the number of groups that survive the velocity filter meanwhile light bars show the number of groups that do not.
Errors are estimated by the standard deviation. 
%
}
\label{fig:optics_survivors}
\end{center}
\end{figure}

 Figure~\ref{fig:optics_survivors} shows the percentage of overdensities that survive the velocity filter as a function of the $\epsilon$ parameter (upper panel). The values were calculated as the mean of the 30 mock catalogues, while the error bars are the standard deviation. Each curve corresponds to a different threshold on the minimum number of members that must contain an overdensity. Even though, the percentage of surviving systems for $N_{mem}=5$ is the lowest, we choose this value with the intention of give higher probability to identify large galaxy systems.  
Bottom panel shows similar information, but in this case, we plot the number of overdensities that survive or not. As can be seen, although the percentage of overdensities that survive with $\epsilon = 0.07$ is the lowest, it is also for this value of $\epsilon$ that the largest number of surviving overdensities are found. We decided to choose this value for the $\epsilon$ parameter in order to extract the largest number of candidates for groups of galaxies and/or clusters, which should be confirmed {\it a posteriori} by means spectroscopic measurements.

\section{Galaxy systems identified in the VVV NIRGC with different methods.}
\label{galaxysystems}

In this section, we present the systems and galaxy members identified with the four methods used in this work.  
Tables~\ref{tab:tab_voro} and \ref{tab:tab_galvoro} are the results of the Voronoi ; 
Tables~\ref{tab:tab_MST} and \ref{tab:tab_galMST} using the MST technique; 
Tables~\ref{tab:tab_optics} and \ref{tab:tab_galoptics} using OPTICS algorithm and 
Tables~\ref{tab:tab_cg} and \ref{tab:tab_galcg} the CGs.
The columns are described in the Notes to the tables.   


\begin{table*}
\caption{Systems identified in the VVV NIRGC using the Voronoi s.}
 \label{tab:tab_voro}
    \centering
    \begin{tabular}{cccccrcccc}
    \hline
    \hline
       Voronoi  & RA & Dec & $l$ & $b$ & $N_{\rm mem}$ &  Radius & $\mu_K$ & Ks$_{\rm b}$ & Ks$_{\rm f}$\\
         ID  & (J2000) & (J2000) & [deg] & [deg] &  & [arcmin] & [mag \, arcsec$^{-2}$]   & [mag]  & [mag]\\
          \hline
        1 &  11:45:35.78 & -60:45:30.11  &     294.974 &       1.092 &   12 &      9.317 &      26.22 &    13.22 &    14.98 \\
        2 &  11:47:50.11 & -59:45:34.31  &     294.995 &       2.129 &    8 &      4.199 &      25.71 &    14.02 &    15.49 \\
        3 &  11:50:30.66 & -60:59:18.88  &     295.610 &       1.014 &   12 &      9.936 &      27.20 &    14.01 &    15.47 \\
        4 &  11:53:53.68 & -61:13:23.25  &     296.062 &       0.879 &    7 &      5.649 &      25.83 &    12.91 &    15.32 \\
        5 &  11:54:36.17 & -61:03:09.00  &     296.108 &       1.064 &    5 &      1.865 &      24.24 &    13.94 &    15.60 \\
        6 &  11:56:57.24 & -60:14:42.00  &     296.216 &       1.914 &    5 &      4.542 &      25.09 &    12.29 &    15.24 \\
        7 &  12:06:01.85 & -63:07:16.95  &     297.826 &      -0.701 &    6 &      3.710 &      25.69 &    13.64 &    15.48 \\
        8 &  12:06:38.32 & -62:37:11.42  &     297.807 &      -0.195 &    7 &      3.999 &      25.43 &    13.54 &    15.25 \\
        9 &  12:10:04.25  & -61:25:39.67 &     298.006 &       1.048 &    6 &      7.520 &      26.98 &    13.84 &    15.28 \\
       10 &  12:10:17.52 & -63:34:06.28  &     298.372 &      -1.062 &    8 &      8.804 &      26.78 &    13.49 &    15.28 \\
        \hline  
          \multicolumn{10}{p{.8\textwidth}}{Notes. The system identification number is shown in column (1); equatorial and Galactic coordinates of the centre in columns (2) to (5); the number of galaxy members in column (6); the estimated radius in arcmin in column (7); the mean $K_{s}$-band surface brightness ($\mu_K$) averaged over that radius in column (8); and the $K_{s}$-band observer-frame extinction-corrected apparent magnitude of the brightest (Ks$_{\rm b}$) and faintest (Ks$_{\rm f}$) galaxy member in columns (9) and (10), respectively. This table is available in electronic form.}
              \end{tabular}
\end{table*}

\begin{table}
\caption{Galaxy members identified in the VVV NIRGC using Voronoi .}
\label{tab:tab_galvoro}
\centering
\setlength{\tabcolsep}{3pt}
\begin{tabular}{ccccc}
\hline
\hline
Voronoi & VVV NIRG & $l$ & $b$ & $K_{s}$\\
ID & ID & [deg] & [deg] & [mag] \\
\hline
1  & VVV-J114420.35-604407.8   &  294.820  &    1.075  & 15.56\\
1  & VVV-J114506.45-604137.8   &  294.900  &    1.139  & 15.60\\
1  & VVV-J114516.92-604110.1   &  294.919  &    1.152  & 15.78\\
1  & VVV-J114520.63-604541.7   &  294.945  &    1.081  & 14.53\\
1  & VVV-J114525.99-604815.4   &  294.967  &    1.043  & 15.28\\
1  & VVV-J114541.97-604746.6   &  294.996  &    1.059  & 15.00\\
1  & VVV-J114547.73-604631.3   &  295.002  &    1.082  & 14.07\\
1  & VVV-J114558.22-603854.6   &  294.991  &    1.210  & 14.33\\
1  & VVV-J114606.45-604138.4   &  295.019  &    1.170  & 14.95\\
1  & VVV-J114608.67-604008.0   &  295.017  &    1.196  & 14.85\\

\hline
\multicolumn{5}{p{.4\textwidth}}{Notes. The Voronoi system identification is shown in column (1); the VVV NIRGC galaxy identification in column (2) and their Galactic coordinates and $K_{s}$ magnitudes in columns (3) to (5). This table is available in electronic form.}
\end{tabular}
\end{table}

\begin{table*}
\caption{Systems identified in the VVV NIRGC using the MST technique.}
 \label{tab:tab_MST}
    \centering
    \begin{tabular}{cccccrcccc}
    \hline
    \hline
    MST & RA & Dec & $l$ & $b$ & $N_{\rm mem}$ &  Radius & $\mu_K$ & Ks$_{\rm b}$ & Ks$_{\rm f}$ \\
    ID  & (J2000) & (J2000) & [deg] & [deg] &  & [arcmin] & [mag \, arcsec$^{-2}$]   & [mag] & [mag]\\
\hline
        1 &  11:48:18.02 & -59:46:51.07 &     295.057 &       2.122 &    5 &      1.452 &      23.90 &    14.02 &    15.34 \\
        2 &  11:54:24.12 & -61:13:18.09 &     296.121 &       0.894 &    6 &      2.192 &      23.75 &    12.91 &    15.32 \\
        3 &  11:54:37.30 & -61:02:01.41 &     296.106 &       1.083 &    6 &      2.183 &      24.48 &    13.94 &    15.60 \\
        4 &  12:05:47.45 & -63:06:39.71 &     297.798 &      -0.695 &    5 &      2.202 &      24.71 &    13.64 &    15.48 \\
        5 &  12:14:17.99 & -61:20:20.23 &     298.493 &       1.212 &    5 &      3.132 &      25.38 &    13.76 &    15.10 \\
        6 &  12:17:19.67 & -63:41:01.82 &     299.162 &      -1.062 &    6 &      1.494 &      23.36 &    13.20 &    15.39 \\
        7 &  12:18:30.59 & -63:47:03.59 &     299.305 &      -1.145 &    5 &      2.347 &      25.32 &    14.85 &    15.43 \\
        8 &  12:19:19.59 & -61:38:51.37 &     299.129 &       0.987 &    5 &      1.702 &      24.42 &    14.63 &    15.35 \\
        9 &  12:19:20.41 & -63:33:24.75 &     299.367 &      -0.907 &    6 &      3.345 &      25.46 &    13.92 &    15.18 \\
       10 &  12:25:41.76 & -60:35:44.20 &     299.775 &       2.118 &    6 &      2.245 &      24.99 &    14.53 &    15.31 \\
\hline
 \multicolumn{10}{p{.8\textwidth}}{Notes. The system identification number is shown in column (1); equatorial and Galactic coordinates of the centre in columns (2) to (5); the number of galaxy members in column (6); the estimated radius in arcmin in column (7); the mean $K_{s}$-band surface brightness averaged over that radius in column (8); and the $K_{s}$-band observer-frame extinction-corrected apparent magnitude of the brightest  and faintest galaxy member in columns (9) and (10), respectively. This table is available in electronic form.}
     \end{tabular}
\end{table*}

\begin{table}
\caption{Galaxy members identified in the VVV NIRGC using  MST technique.}
\label{tab:tab_galMST}
\centering
\setlength{\tabcolsep}{3pt}
\begin{tabular}{ccccc}
\hline
\hline
MST & VVV NIRG & $l$   &   $b$   & $K_{s}$\\
ID  & ID       & [deg] & [deg] & [mag] \\
\hline
1  & VVV-J114810.25-594546.7  &    295.037  & 2.135  & 16.44 \\
1  & VVV-J114815.25-594815.6  &    295.058  & 2.098  & 17.13 \\
1  & VVV-J114817.62-594756.8  &    295.061  & 2.104  & 16.24 \\
1  & VVV-J114817.82-594746.0  &    295.061  & 2.107  & 15.50 \\
1  & VVV-J114823.46-594534.2  &    295.063  & 2.146  & 15.60 \\
2  & VVV-J115408.19-611422.0  &    296.094  & 0.869  & 15.61 \\
2  & VVV-J115420.90-611526.8  &    296.123  & 0.857  & 14.64 \\
2  & VVV-J115428.00-611507.3  &    296.135  & 0.866  & 13.98 \\
2  & VVV-J115437.55-611431.1  &    296.152  & 0.880  & 14.33 \\
2  & VVV-J115440.03-611214.1  &    296.148  & 0.918  & 15.23 \\
\hline
\multicolumn{5}{p{.4\textwidth}}{Notes. The MST system identification is shown in column (1); the VVV NIRGC galaxy identification in column (2) and their Galactic coordinates and $K_{s}$ magnitudes in columns (3) to (5). This table is available in electronic form.}
\end{tabular}
\end{table}

\begin{table*}
\caption{Systems identified in the VVV NIRGC using the OPTICS algorithm.}
\label{tab:tab_optics}
\centering
\begin{tabular}{cccccrcccc}
\hline
\hline
OPTICS & RA & Dec & $l$ & $b$ & $N_{\rm mem}$ & Radius & $\mu_K$ & Ks$_{\rm b}$ & Ks$_{\rm f}$ \\
ID & (J2000) & (J2000) & [deg] & [deg] &  & [arcmin] & [mag \, arcsec$^{-2}$]   & [mag] & [mag]\\
\hline
1	& 11:45:54.53	& -60:47:57.94	& 295.022	& 1.062	    &  7	& 4.721	& 25.00	& 12.82	& 14.74\\
2	& 11:48:09.11	& -59:47:16.53	& 295.041	& 2.111	    &  8	& 4.219	& 25.70	& 14.02	& 15.34\\
3	& 11:51:49.16	& -61:38:54.24	& 295.916	& 0.409	    &  5	& 3.094	& 25.20	& 13.36	& 15.19\\
4	& 11:54:40.70	& -61:14:16.62	& 296.157	& 0.885	    &  7	& 3.912	& 24.93	& 12.91	& 15.32\\
5	& 11:54:49.97	& -61:03:21.85	& 296.136	& 1.067	    &  7	& 3.549	& 25.36	& 13.94	& 15.60\\
6	& 11:57:02.08	& -60:12:29.09	& 296.218	& 1.953	    &  5	& 3.347	& 25.23	& 13.53	& 15.24\\
7	& 12:06:01.85	& -63:07:16.94	& 297.826	& -0.701    &  7	& 3.710	& 25.47	& 13.64	& 15.48\\
8	& 12:06:22.76	& -63:34:17.52	& 297.943	& -1.137    &  9	& 7.082	& 26.09	& 12.83	& 15.10\\
9	& 12:06:30.82	& -62:37:16.66	& 297.793	& -0.199	& 10	& 5.210	& 25.15	& 12.38	& 15.25\\
10	& 12:08:37.15	& -60:25:48.17	& 297.669	& 2.003	    &  5	& 4.173	& 25.78	& 13.79	& 15.50\\

          \hline
          \multicolumn{10}{p{.8\textwidth}}{Notes. The system identification number is shown in column (1); equatorial and Galactic coordinates of the centre in columns (2) to (5); the number of galaxy members in column (6); the estimated radius in arcmin in column (7); the mean $K_{s}$-band surface brightness averaged over that radius in column (8); and the $K_{s}$-band observer-frame extinction-corrected apparent magnitude of the brightest  and faintest galaxy member in columns (9) and (10), respectively. This table is available in electronic form.}
    \end{tabular}
\end{table*}

\begin{table}
\caption{Galaxy members identified in the VVV NIRGC using OPTICS algorithm.}
\label{tab:tab_galoptics}
\centering
\setlength{\tabcolsep}{3pt}
\begin{tabular}{ccccc}
\hline
\hline
OPTICS & VVV NIRG & $l$   & $b$ & $K_{s}$\\
ID     &     ID   & [deg] & [deg] & [mag] \\
\hline
 1 & VVV-J114520.63-604541.7 & 294.945 & 1.081 & 13.46\\
 1 & VVV-J114525.99-604815.4 & 294.966 & 1.043 & 14.26\\
 1 & VVV-J114541.97-604746.6 & 294.996 & 1.058 & 13.55\\
 1 & VVV-J114547.73-604631.3 & 295.002 & 1.082 & 13.27\\
 1 & VVV-J114550.32-605133.2 & 295.028 & 1.002 & 14.46\\
 1 & VVV-J114615.00-604750.5 & 295.061 & 1.074 & 14.74\\
 1 & VVV-J114628.52-605013.7 & 295.097 & 1.043 & 12.82\\
 2 & VVV-J114755.99-595109.5 & 295.030 & 2.041 & 15.18\\
 2 & VVV-J114810.15-595129.6 & 295.060 & 2.043 & 14.23\\
 2 & VVV-J114810.25-594546.7 & 295.037 & 2.135 & 15.28\\

\hline
\multicolumn{5}{p{.4\textwidth}}{Notes. OPTICS system identification is shown in column (1); the VVV NIRGC galaxy identification in column (2) and their Galactic coordinates and $K_{s}$ magnitudes in columns (3) to (5). This table is available in electronic form.}
\end{tabular}
\end{table}

\begin{table*}
\caption{CGs identified in the VVV NIRGC. }
\label{tab:tab_cg}
\centering
\begin{tabular}{cccccrcccc}
\hline
\hline
       CG & RA & Dec & $l$ & $b$ & $N_{\rm mem}$ &  Radius & $\mu_K$ & Ks$_{\rm b}$ & Ks$_{\rm max}$ \\
    ID  & (J2000) & (J2000) & [deg] & [deg] &  & [arcmin] & [mag \, arcsec$^{-2}$]   & [mag] & [mag]\\

\hline
 1 &  13:40:03.33  & -60:27:20.85 &     308.894 &       \ 1.852 &    3 &      1.241 &      21.84 &    11.56 &    15.31 \\
 2 &  13:44:18.15 & -60:20:34.84 &     309.431 &       \ 1.860 &    3 &      3.036 &      23.46 &    11.08 &    14.91 \\
 3 &  14:39:31.62 & -57:44:07.85  &     316.978 &       \ 2.158 &    3 &      3.172 &      23.44 &    11.09 &    14.92 \\
 4 &  15:01:26.71  & -57:24:07.75  &     319.762 &     \ 1.159 &    3 &      1.459 &      21.59 &    11.46 &    14.96 \\
 5 &  11:42:59.27 & -63:50:57.65 &     295.476 &      -1.973 &    3 &      1.167 &      23.11 &    13.02 &    15.24 \\
 6 &  11:48:16.53 & -59:48:00.81  &     295.059 &      \ 2.103 &    3 &      0.295 &      20.85 &    14.02 &    15.34 \\
 7 &  11:50:15.61 & -59:48:39.06 &     295.304 &      \ 2.152 &    3 &      1.534 &      23.40 &    12.97 &    14.70 \\
 8 &  11:53:46.72 & -60:25:31.74 &     295.871 &      \ 1.653 &    3 &      0.761 &      22.82 &    13.91 &    15.43 \\
 9 &  11:54:11.55 & -60:50:57.88 &     296.015 &      \ 1.251 &    3 &      1.291 &      23.53 &    13.35 &    15.36 \\
10 &  11:54:48.21 & -61:02:57.00  &     296.131 &       \ 1.073 &    3 &      0.579 &      22.36 &    13.94 &    15.60 \\
\hline

\multicolumn{10}{p{.85\textwidth}}{Notes. The first four rows correspond to the \emph{restricted} CGs followed by the first 6 \emph{unrestricted} CGs. 
The CG  identification number is shown in column (1); equatorial and Galactic coordinates of the centre in columns (2) to (5); the number of galaxy members in column (6); the angular radius of the smallest circumscribed circle in column (7); the mean $K_{s}$-band surface brightness averaged over that radius in column (8); 
the $K_{s}$-band 
 observer-frame extinction-corrected apparent magnitude of the group brightest galaxy in column (9), and maximum extinction-corrected apparent magnitude at the position of the brightest galaxy in column (10). This table is available in electronic form.}
\end{tabular}
\end{table*}

\begin{table}
\caption{Galaxy members in CGs identified in the VVV NIRGC}
\label{tab:tab_galcg}
\centering
\setlength{\tabcolsep}{3pt}
\begin{tabular}{ccccc}
\hline
\hline
CG     & VVV NIRG & $l$   & $b$ & $K_{s}$\\
ID     &     ID   & [deg] & [deg] & [mag] \\
\hline
1 & VVV-J133959.99-602610.6 &     308.891 &       1.872 &    13.30 \\
1 & VVV-J134003.02-602835.3 &     308.890 &       1.832 &    11.56 \\
1 & VVV-J134012.21-602755.9 &     308.910 &       1.839 &    13.64 \\
2 & VVV-J134403.55-601935.5 &     309.405 &       1.882 &    11.08 \\
2 & VVV-J134413.08-602333.1 &     309.411 &       1.814 &    13.92 \\
2 & VVV-J134423.19-601736.6 &     309.451 &       1.906 &    13.74 \\
3 & VVV-J143911.27-574229.6 &     316.948 &       2.201 &    11.09 \\
3 & VVV-J143933.63-574717.5 &     316.961 &       2.108 &    13.37 \\
3 & VVV-J143934.82-574059.3 &     317.006 &       2.203 &    12.98 \\
4 & VVV-J150116.26-572345.0 &     319.745 &       1.175 &    11.46 \\
\hline
\multicolumn{5}{p{.4\textwidth}}{Notes. The CG identification is shown in column (1); the VVV NIRGC galaxy identification in column (2) and their Galactic coordinates and $K_{s}$ magnitudes in columns (3) to (5). This table is available in electronic form.} 
\end{tabular}
\end{table}


\begin{figure*}
    \centering
    \includegraphics[width=0.24\textwidth]{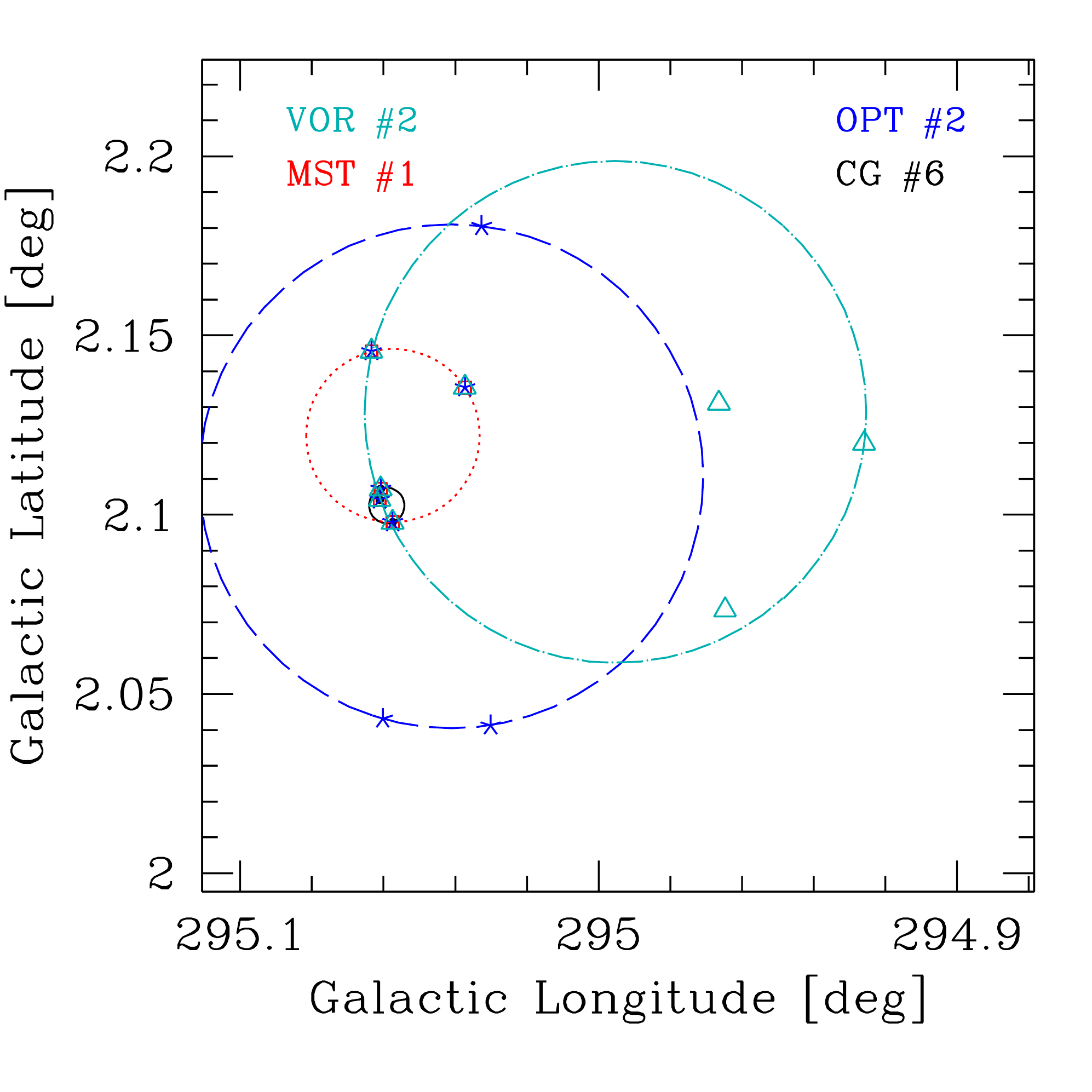}
    \includegraphics[width=0.24\textwidth]{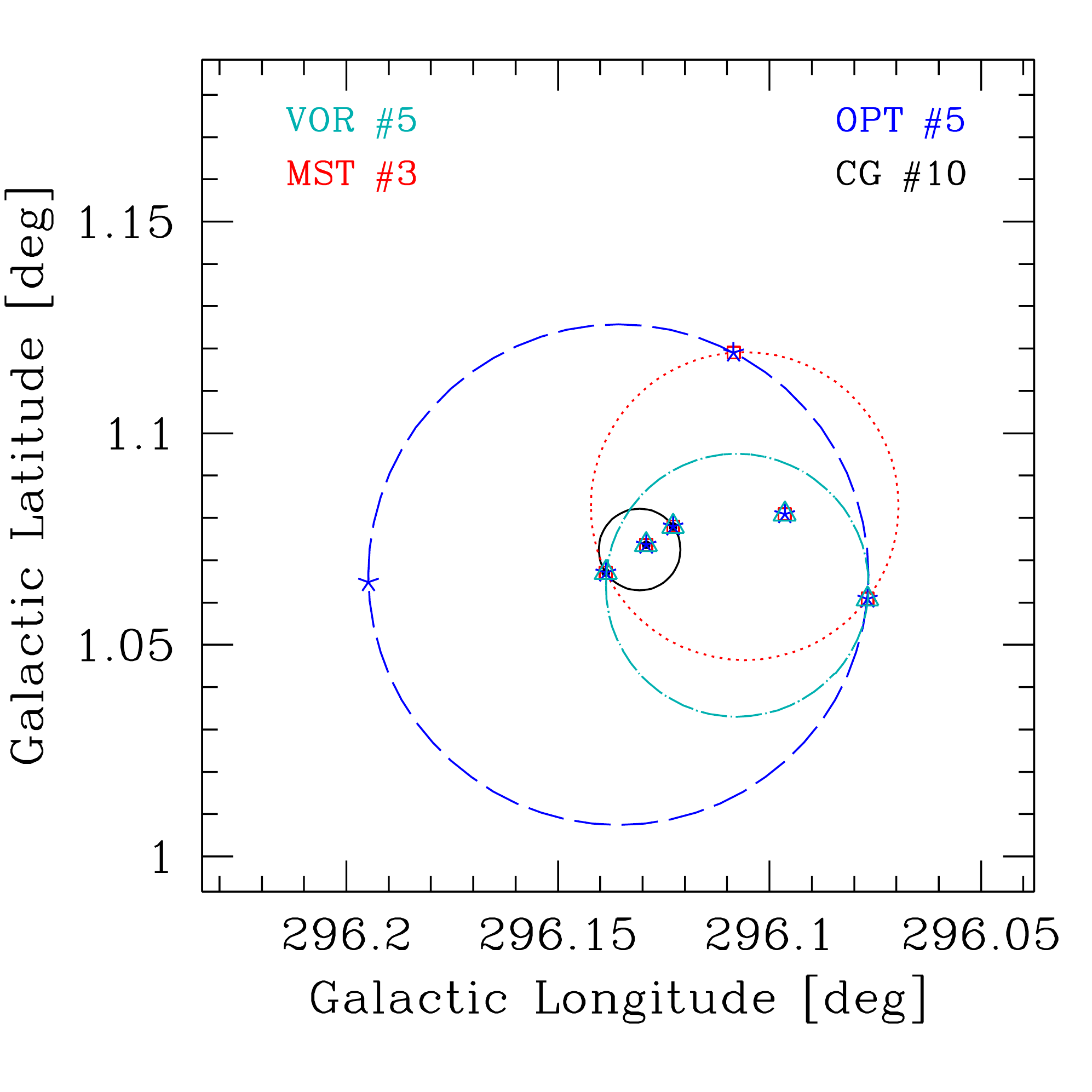}
    \includegraphics[width=0.24\textwidth]{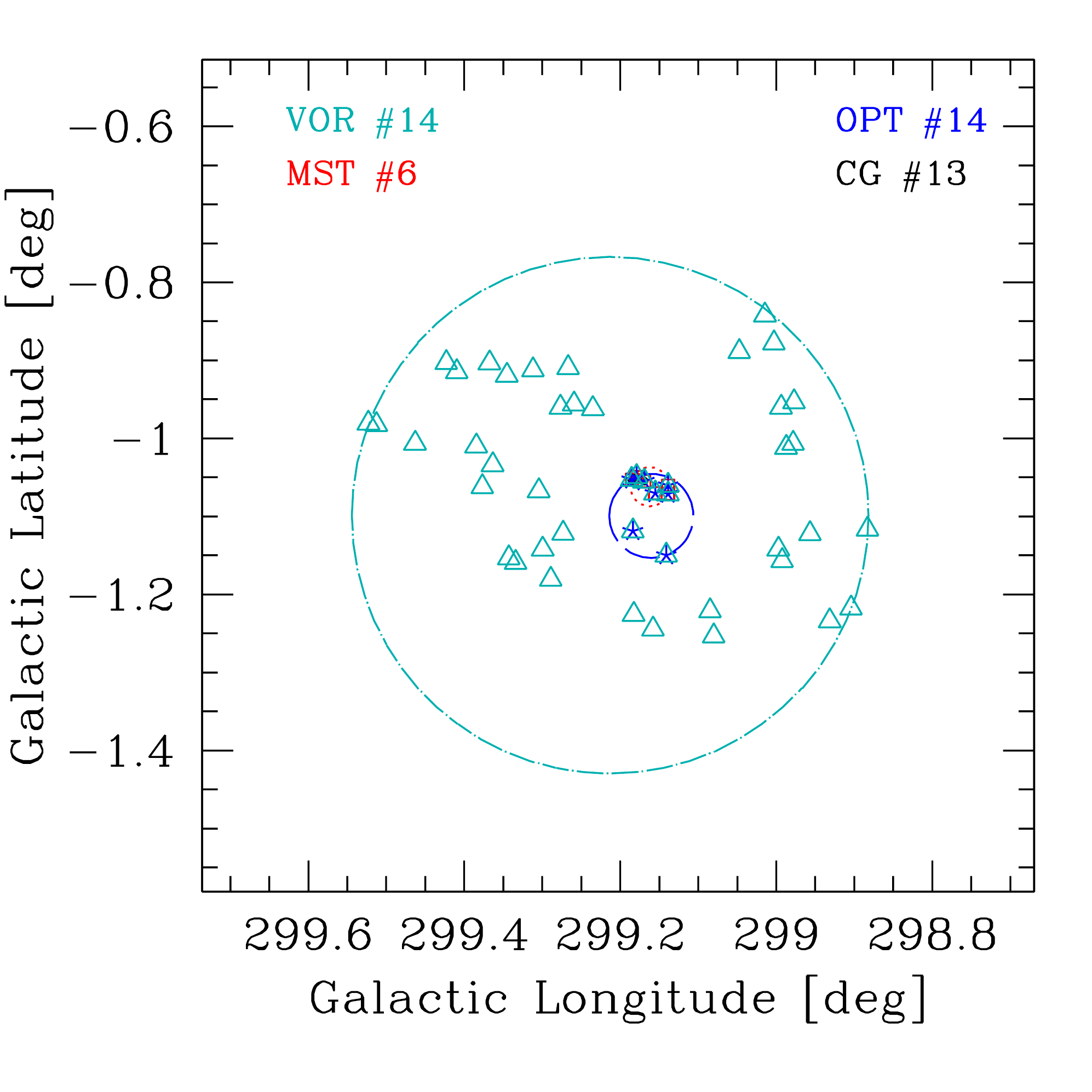}
    \includegraphics[width=0.24\textwidth]{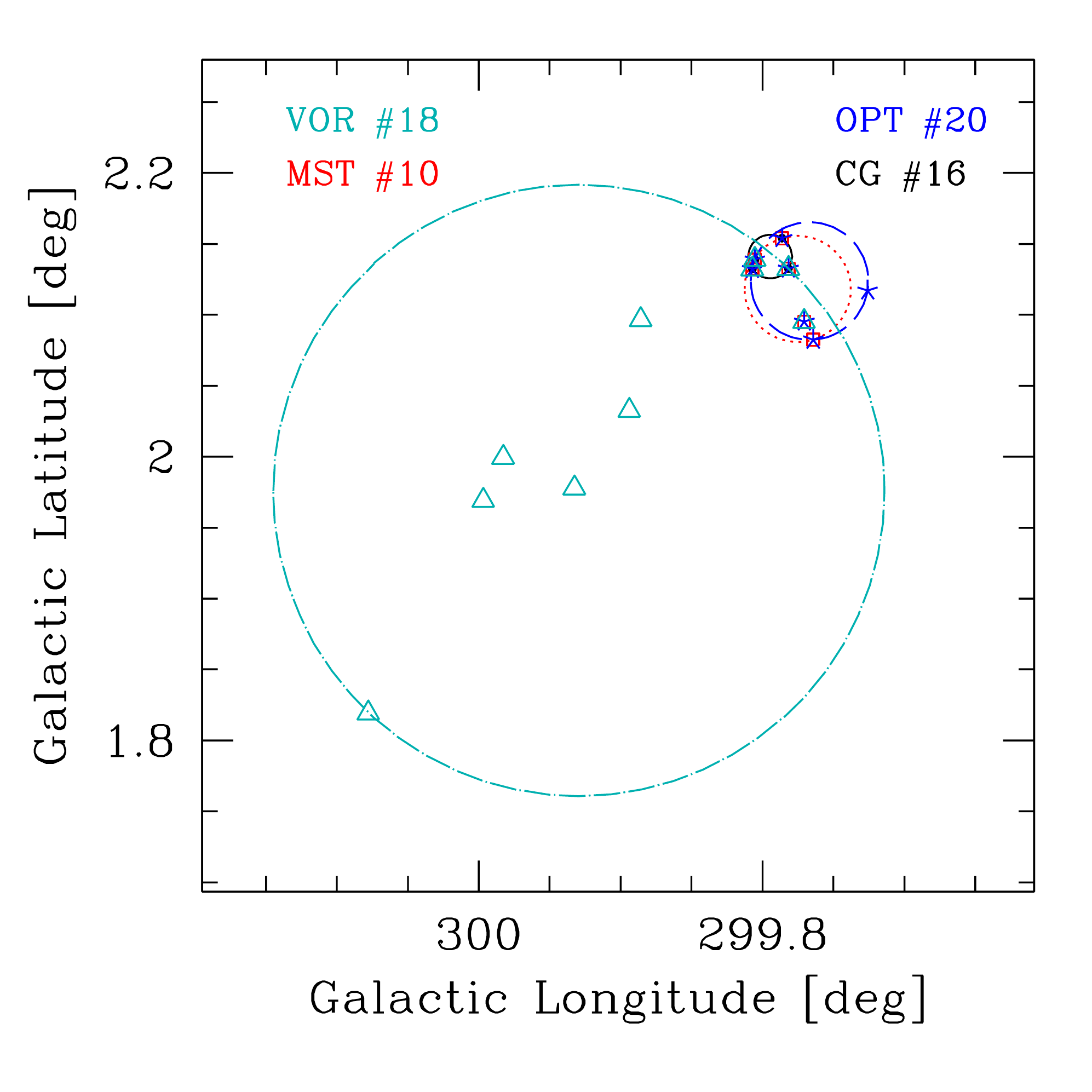}
    \includegraphics[width=0.24\textwidth]{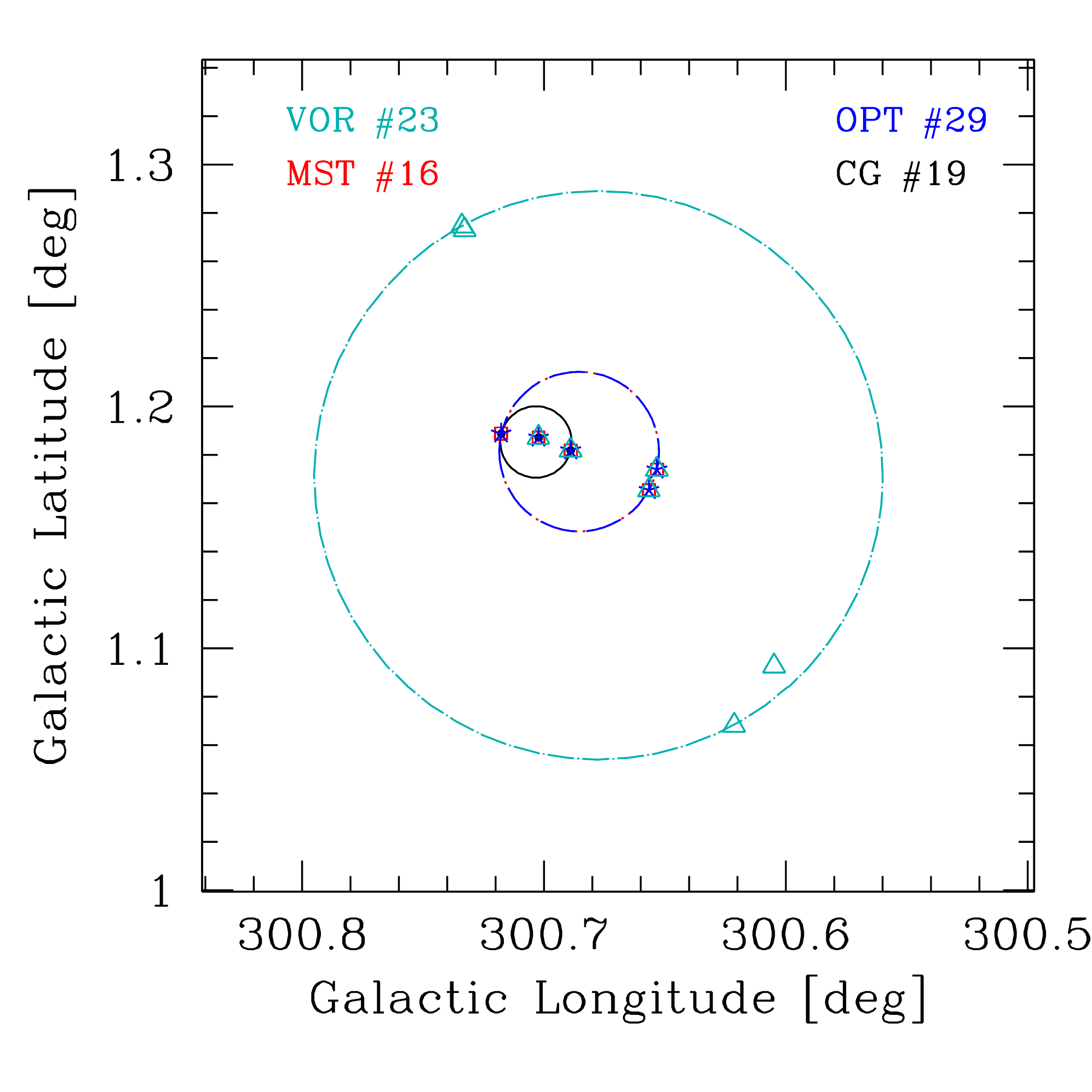}
    \includegraphics[width=0.24\textwidth]{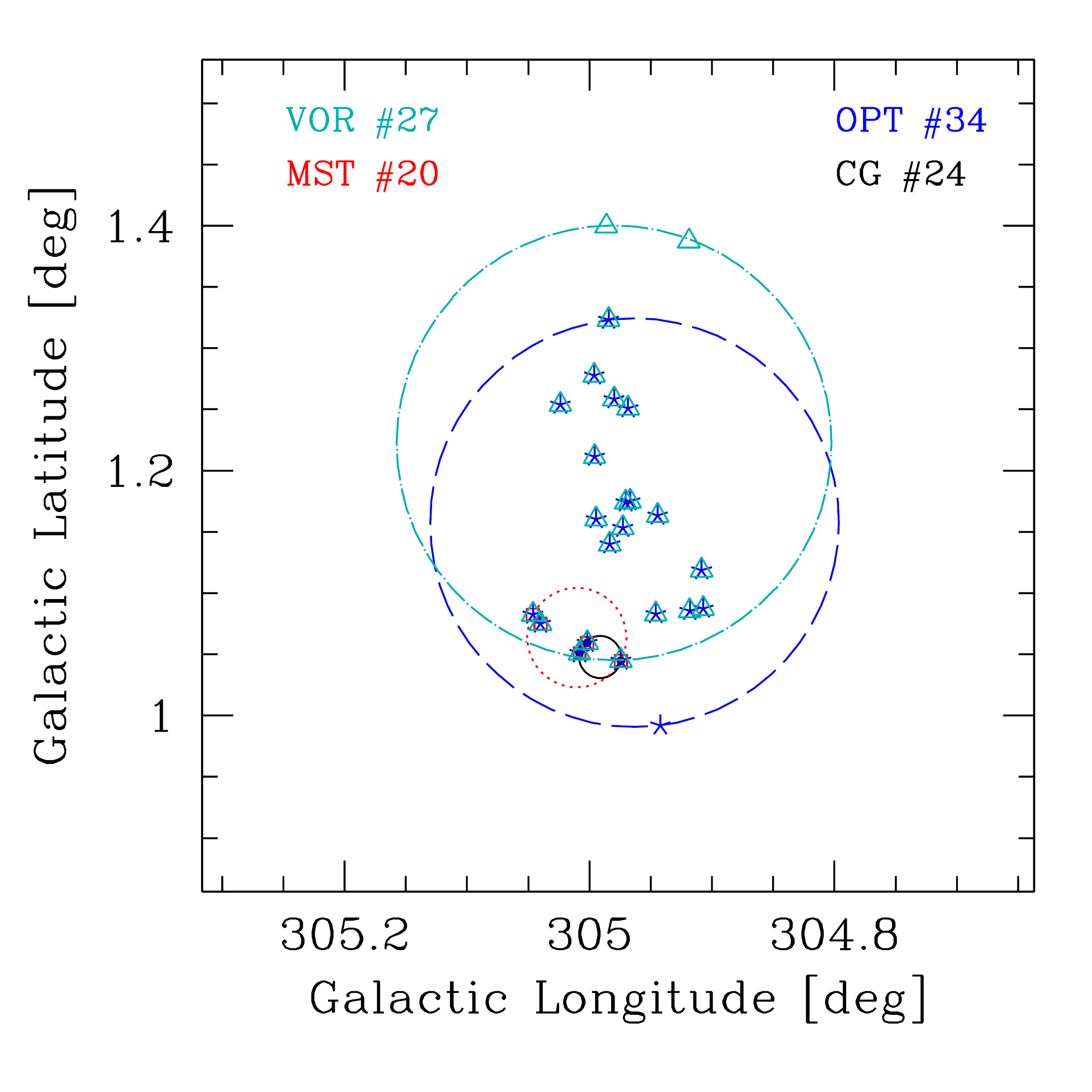}
    \includegraphics[width=0.24\textwidth]{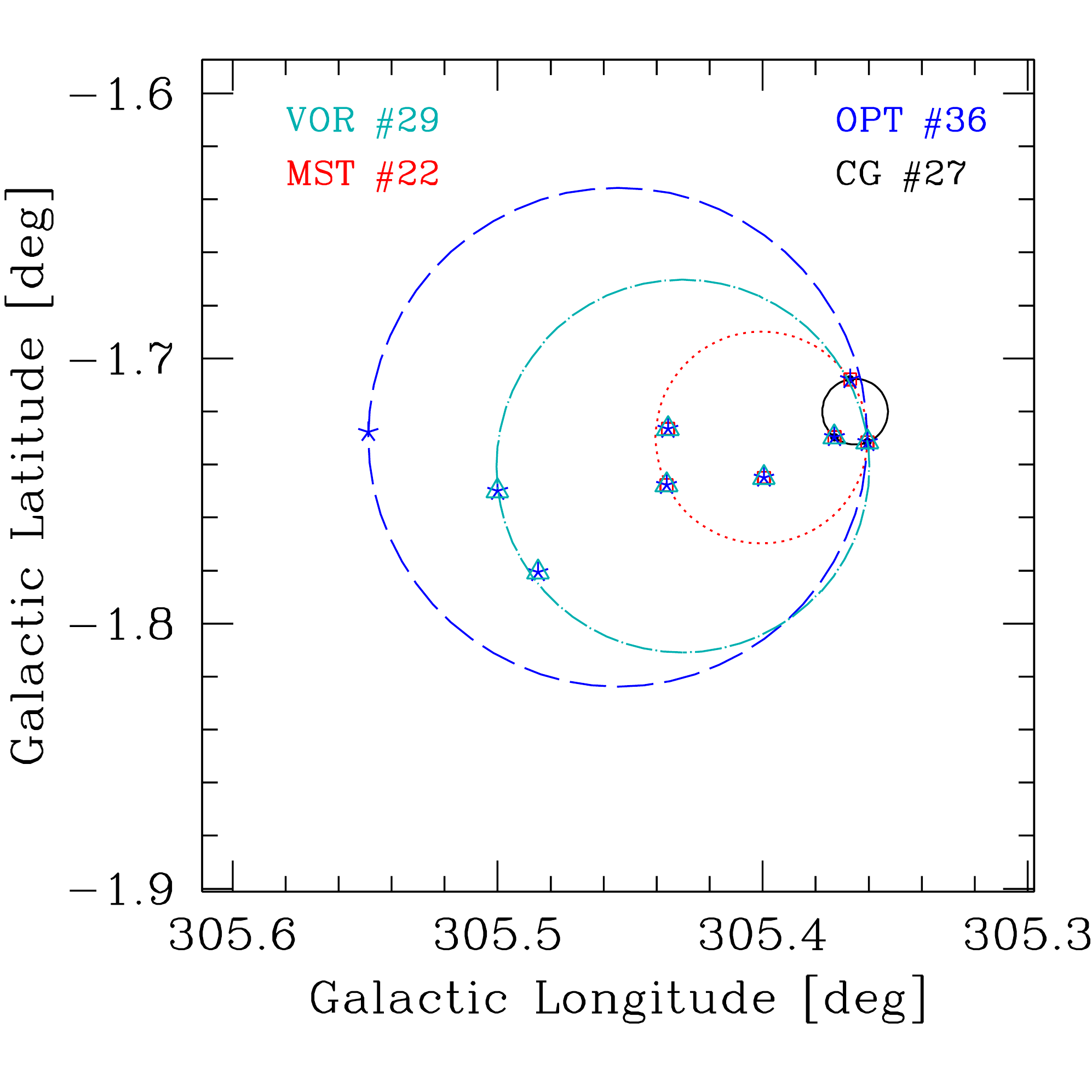}
    \includegraphics[width=0.24\textwidth]{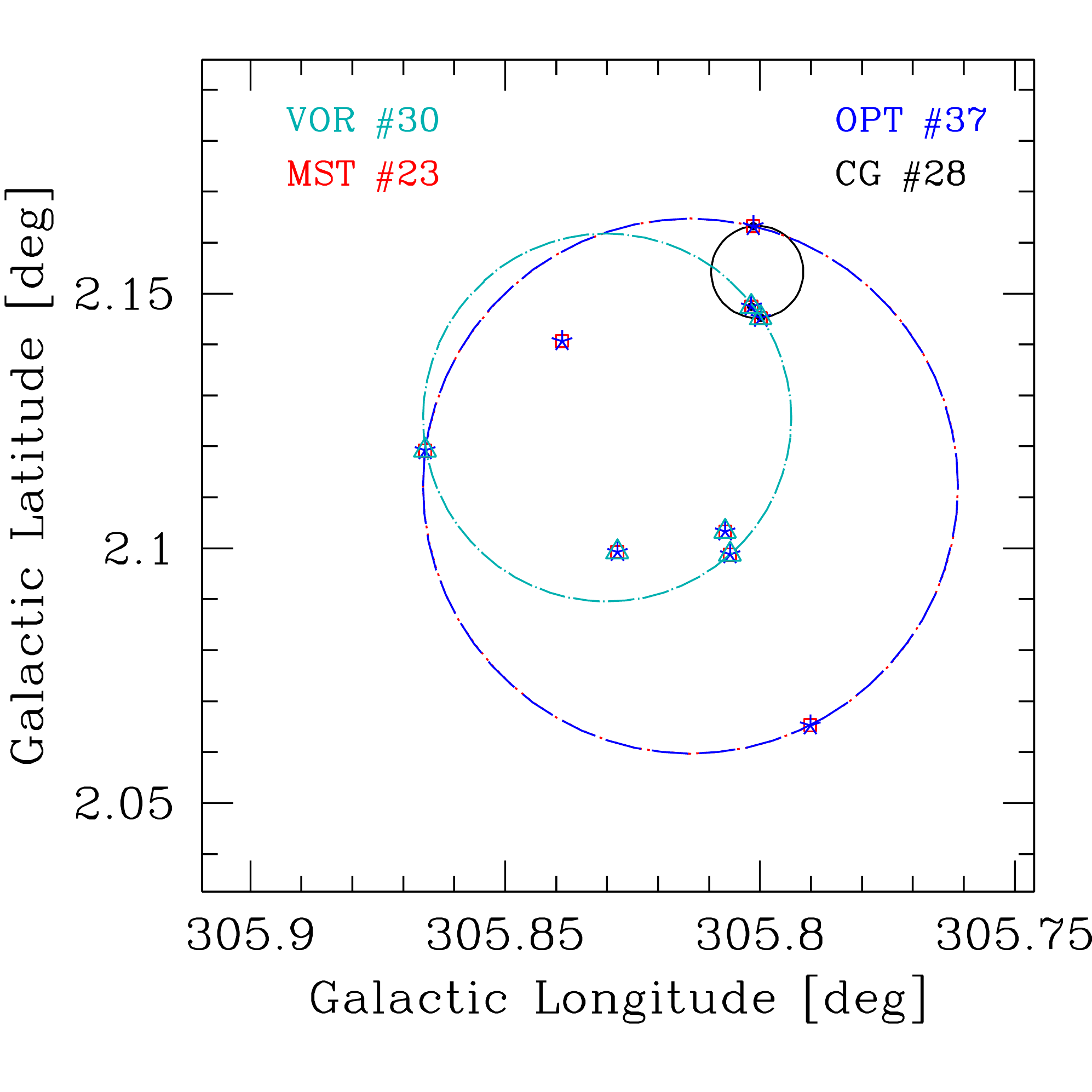}
    \includegraphics[width=0.24\textwidth]{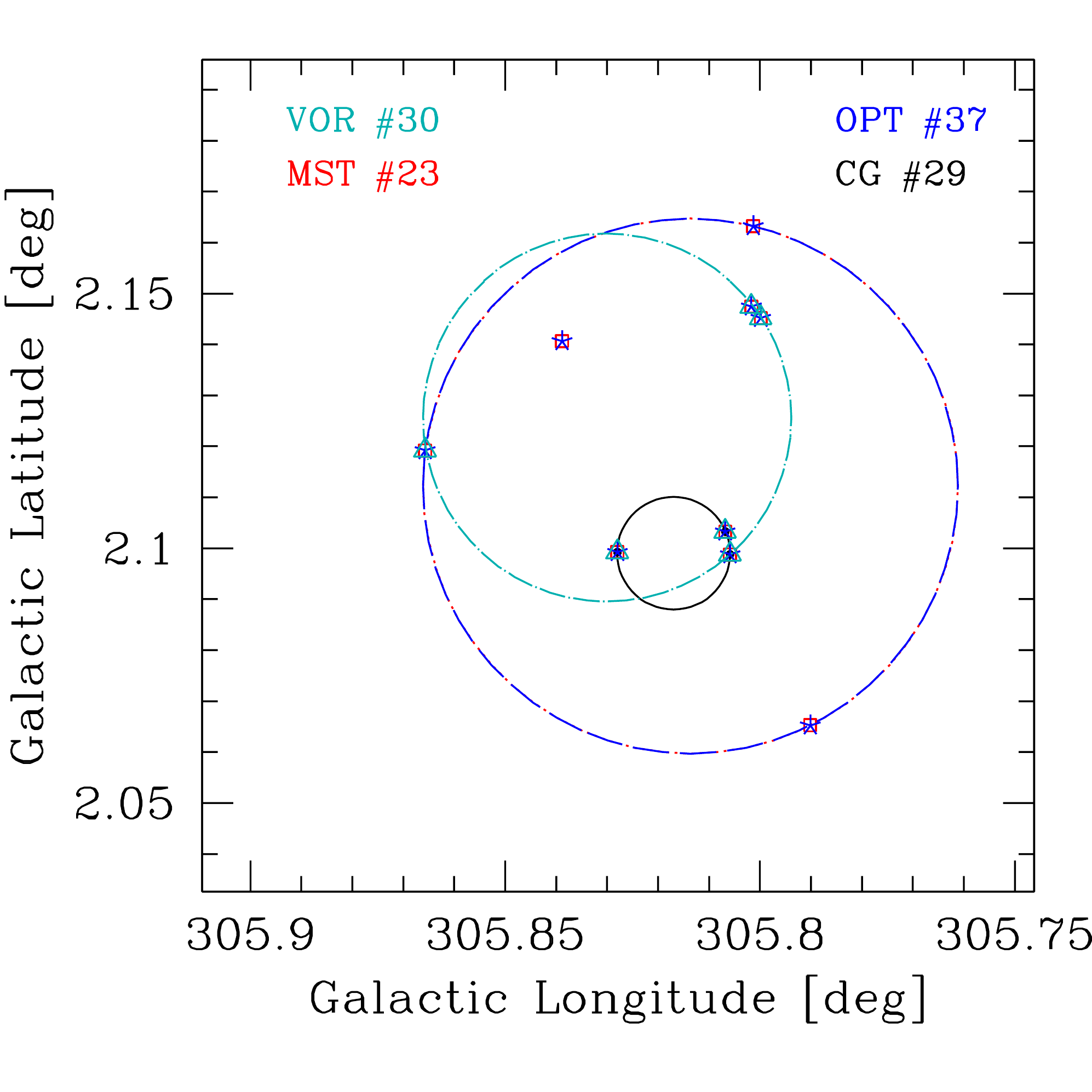}
    \includegraphics[width=0.24\textwidth]{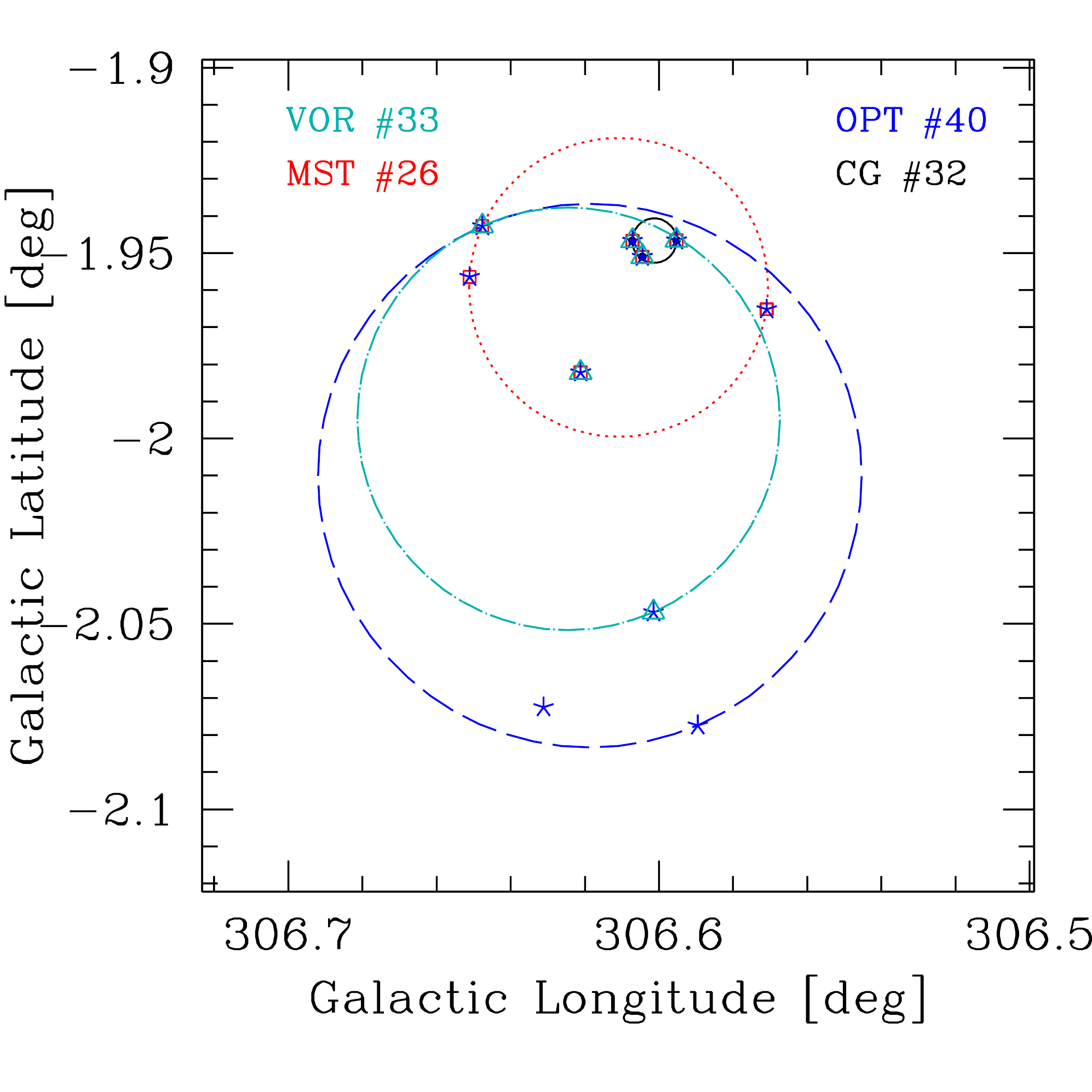}
    \includegraphics[width=0.24\textwidth]{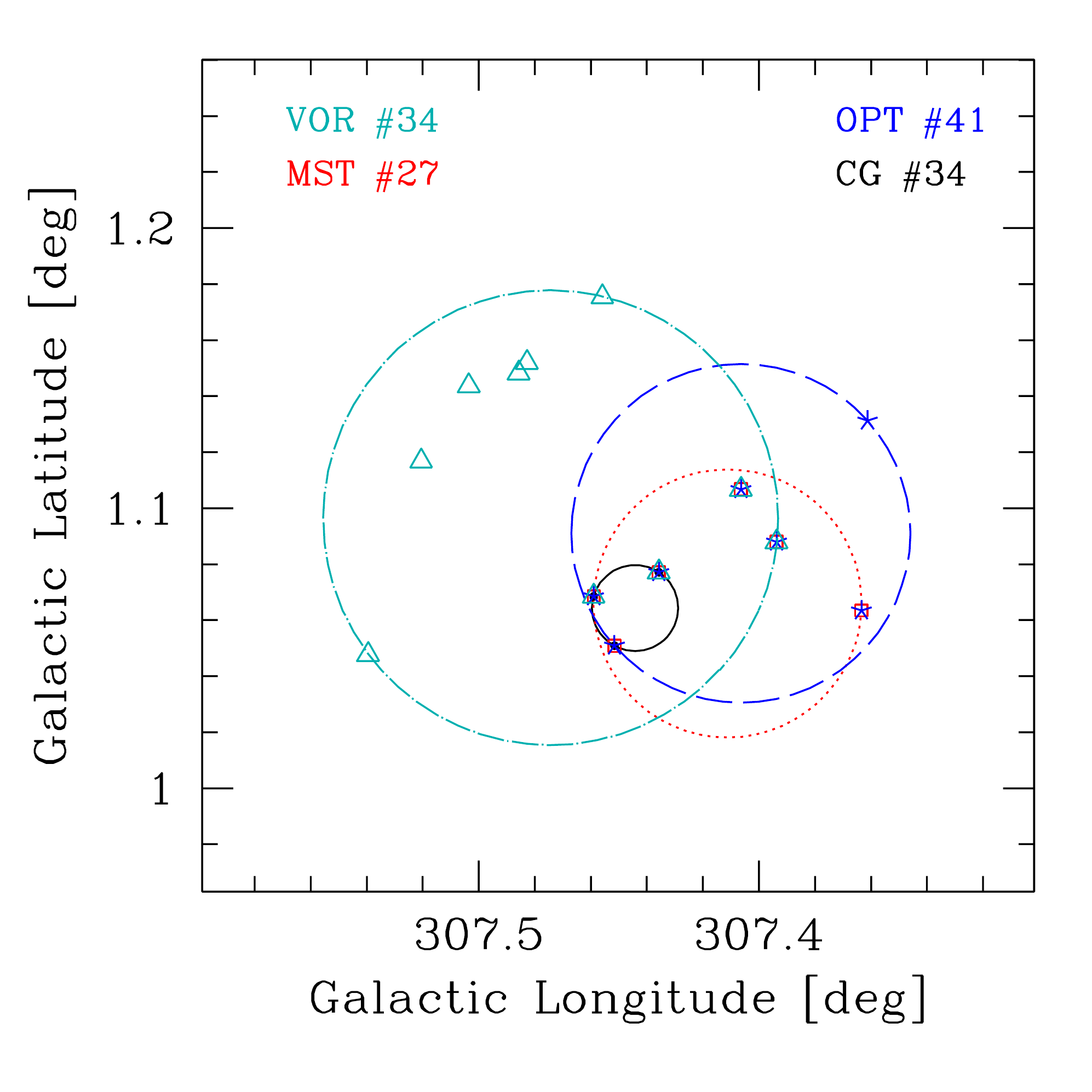}
    \includegraphics[width=0.24\textwidth]{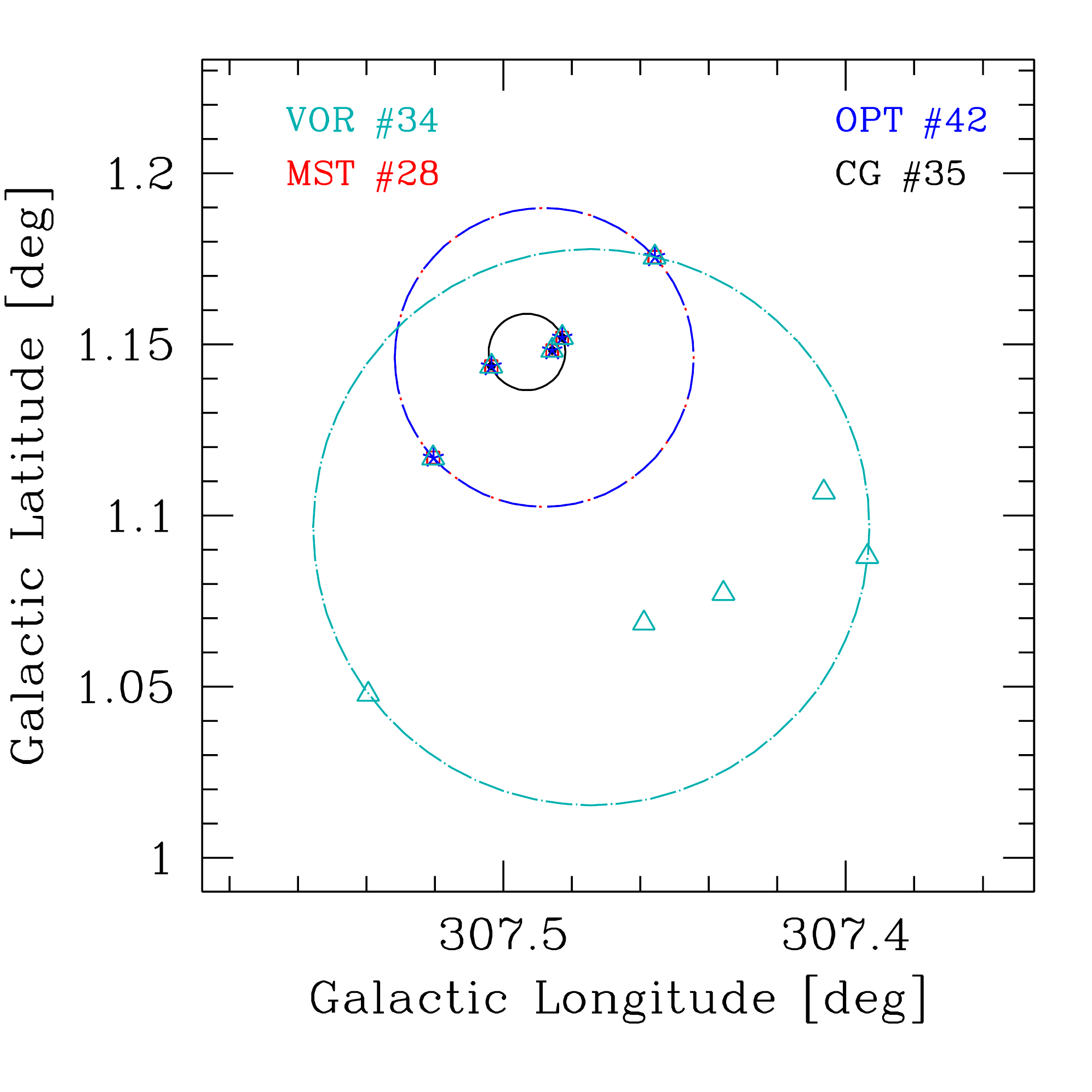}
    \includegraphics[width=0.24\textwidth]{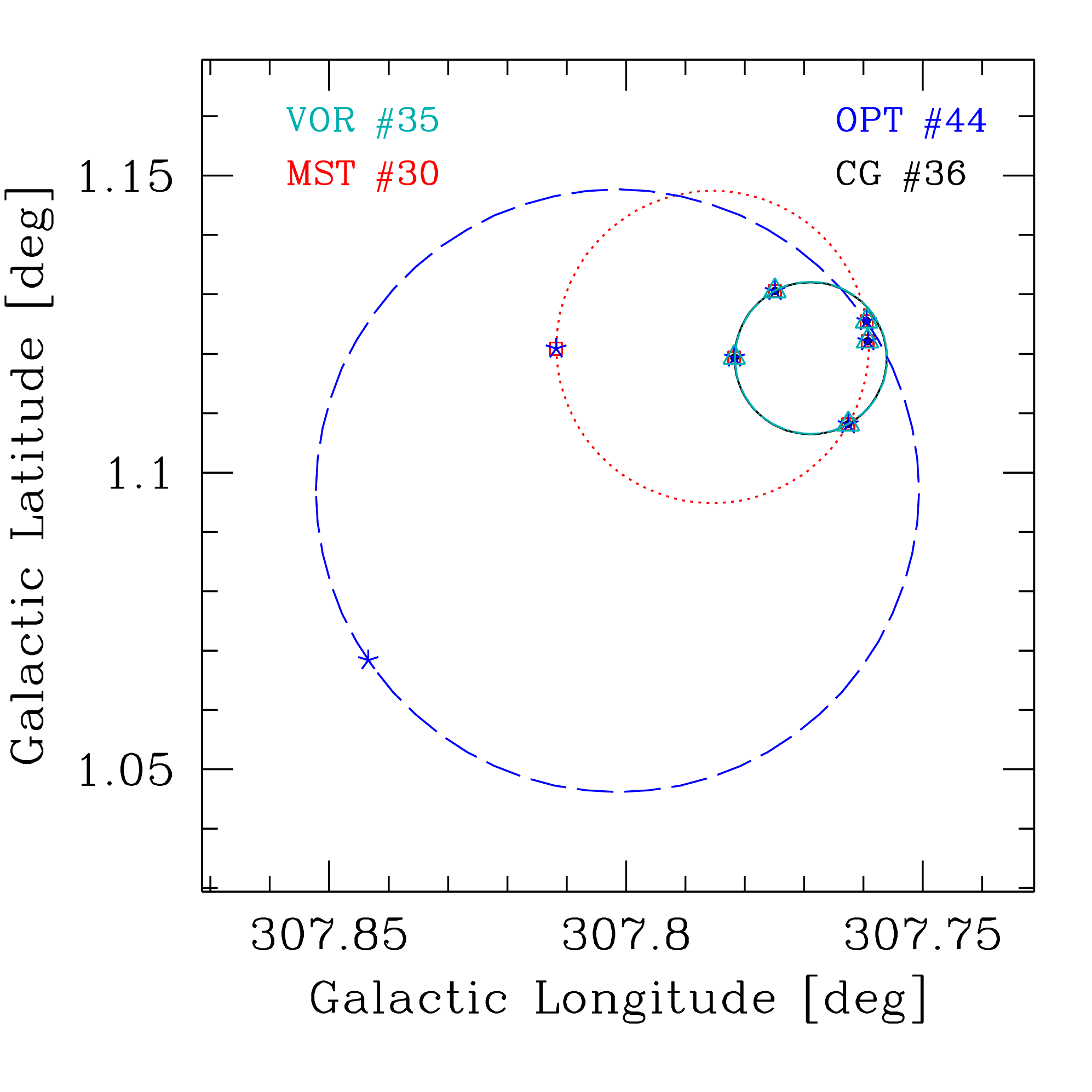}
    \includegraphics[width=0.24\textwidth]{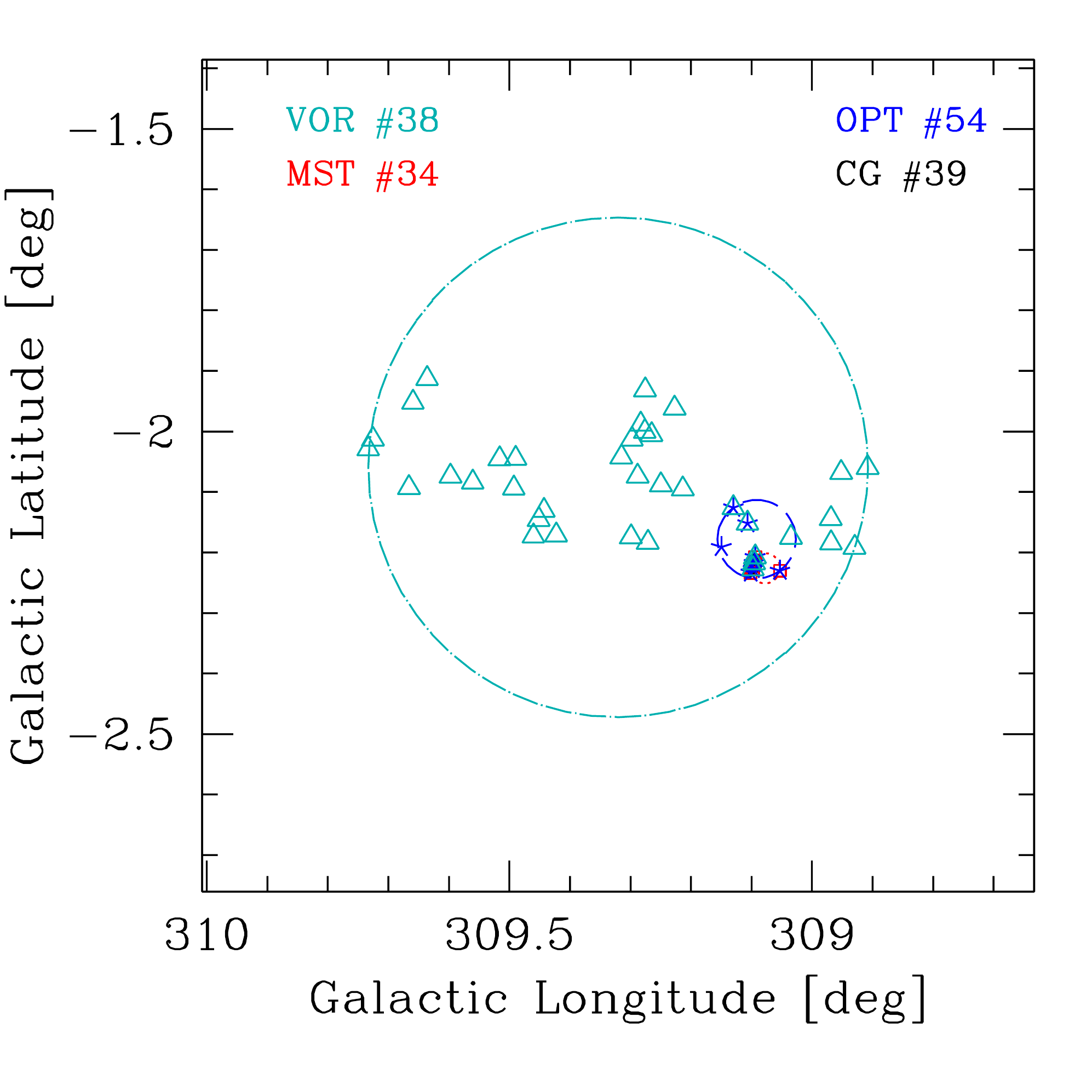}
    \includegraphics[width=0.24\textwidth]{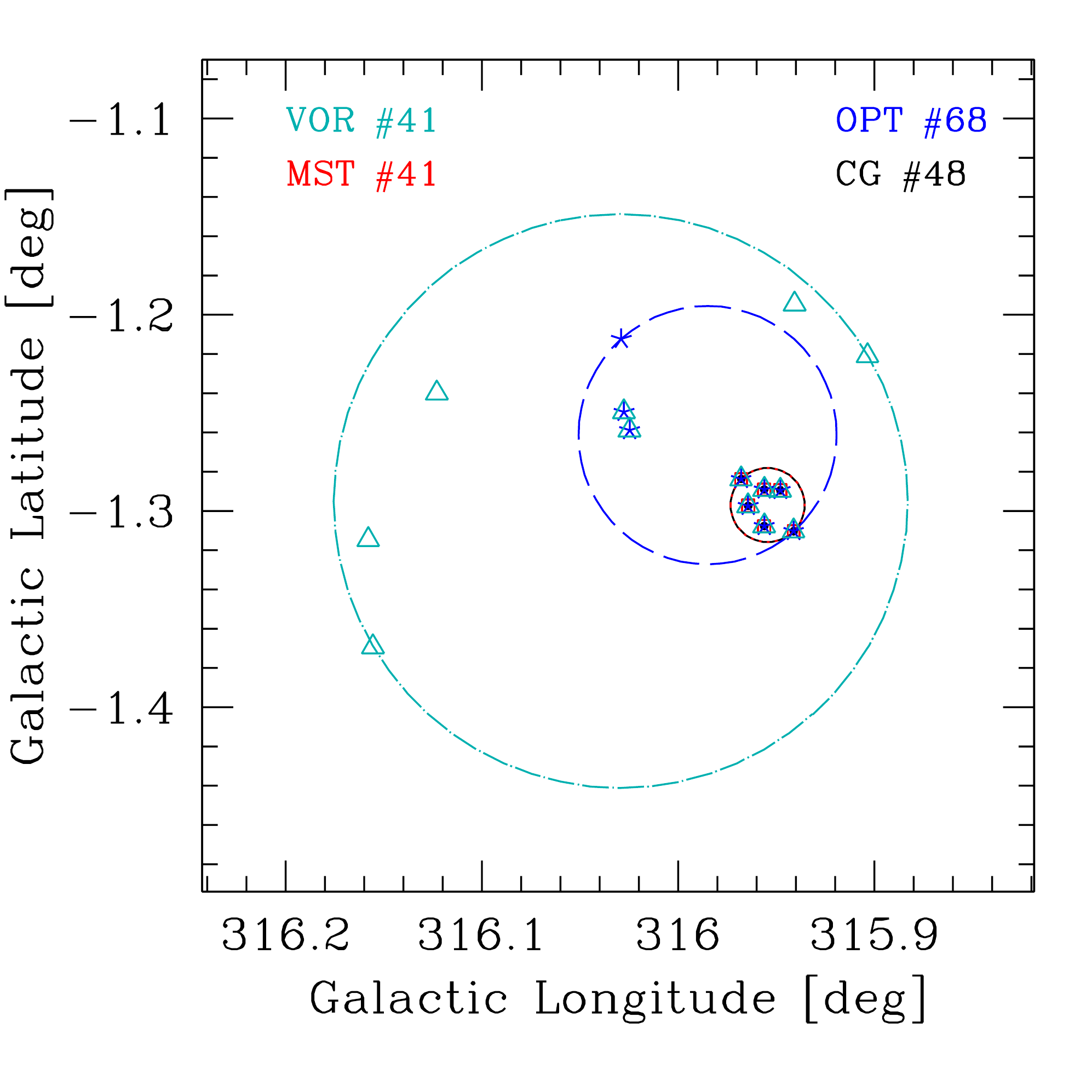}
    \includegraphics[width=0.24\textwidth]{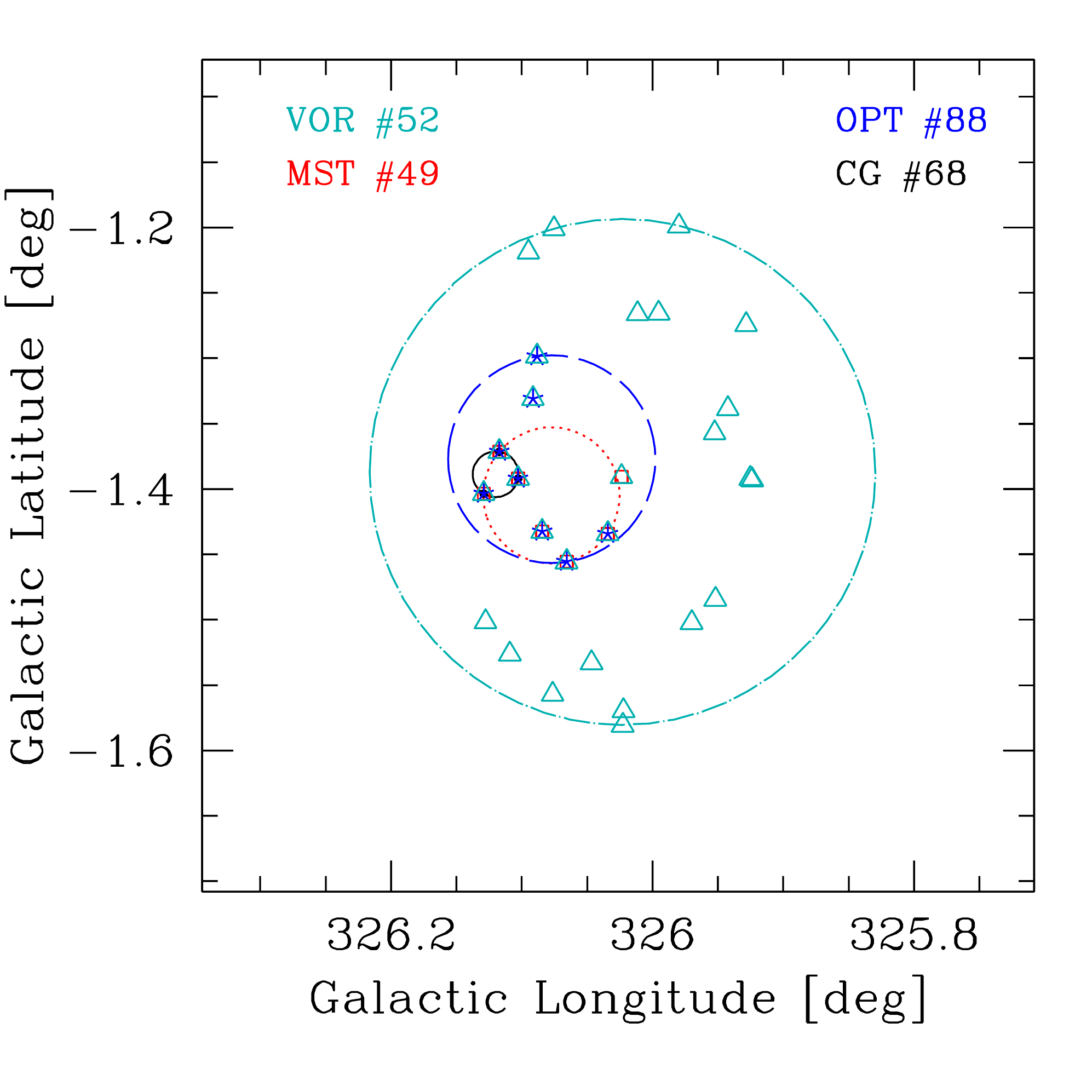}
    \includegraphics[width=0.24\textwidth]{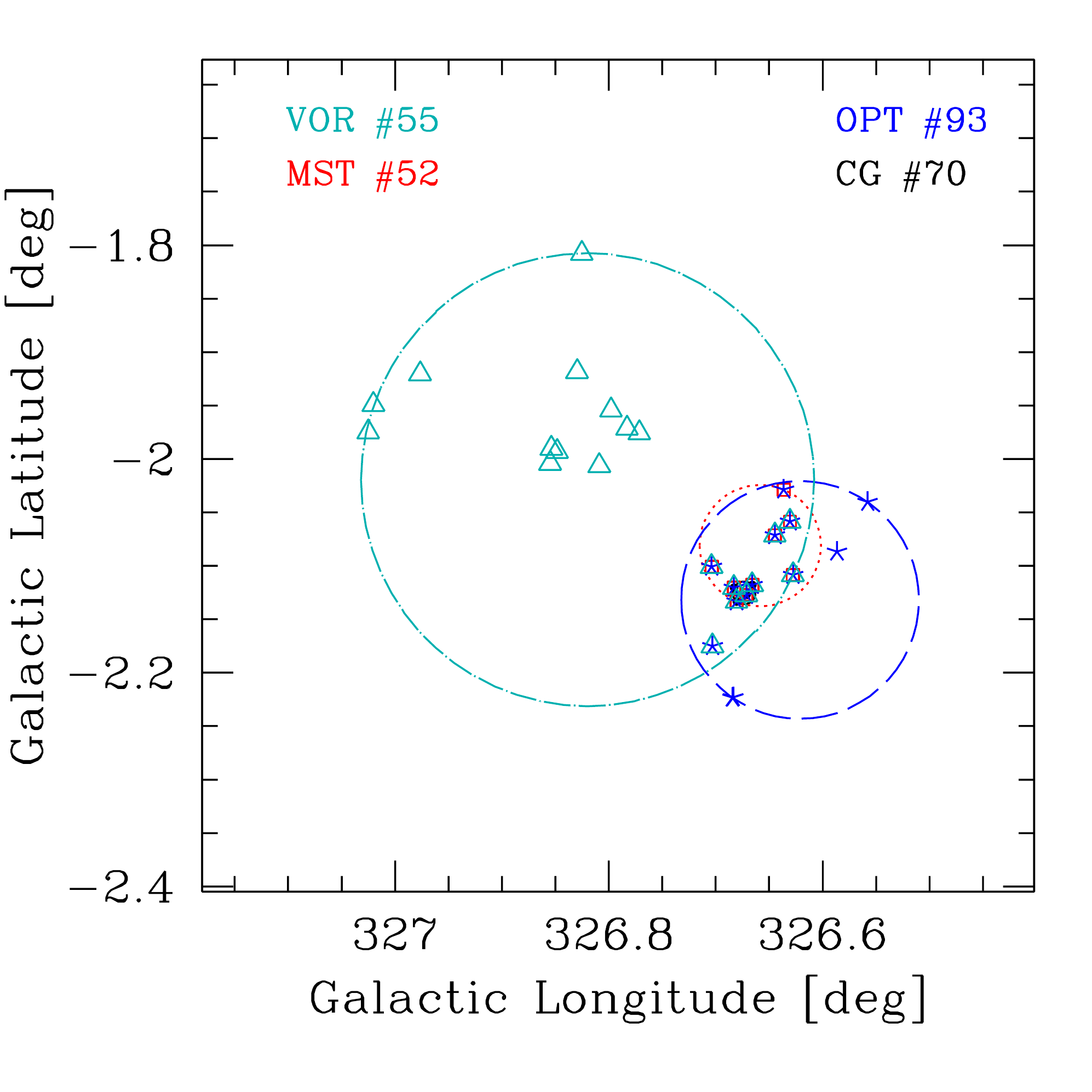}
    \includegraphics[width=0.24\textwidth]{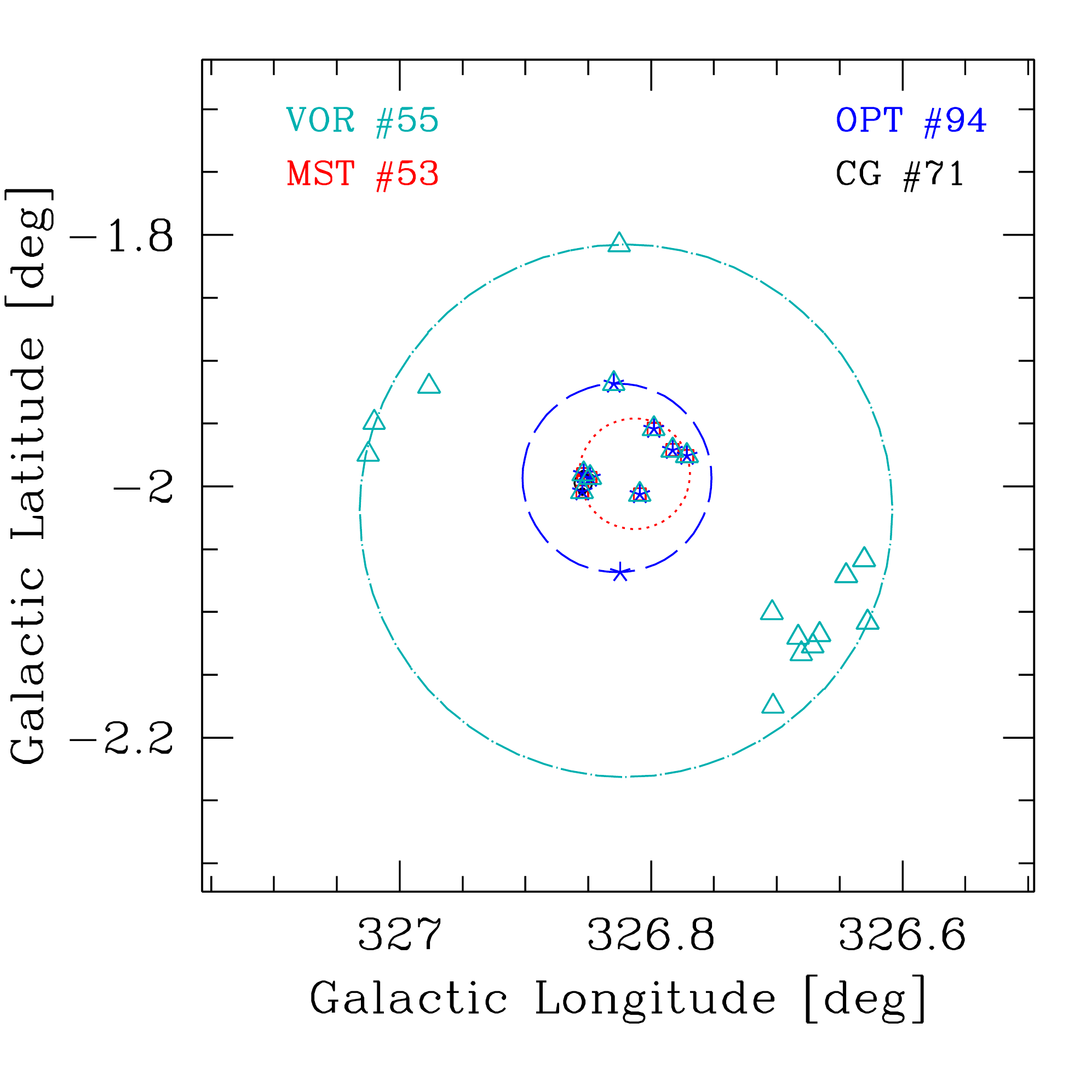}
    \includegraphics[width=0.24\textwidth]{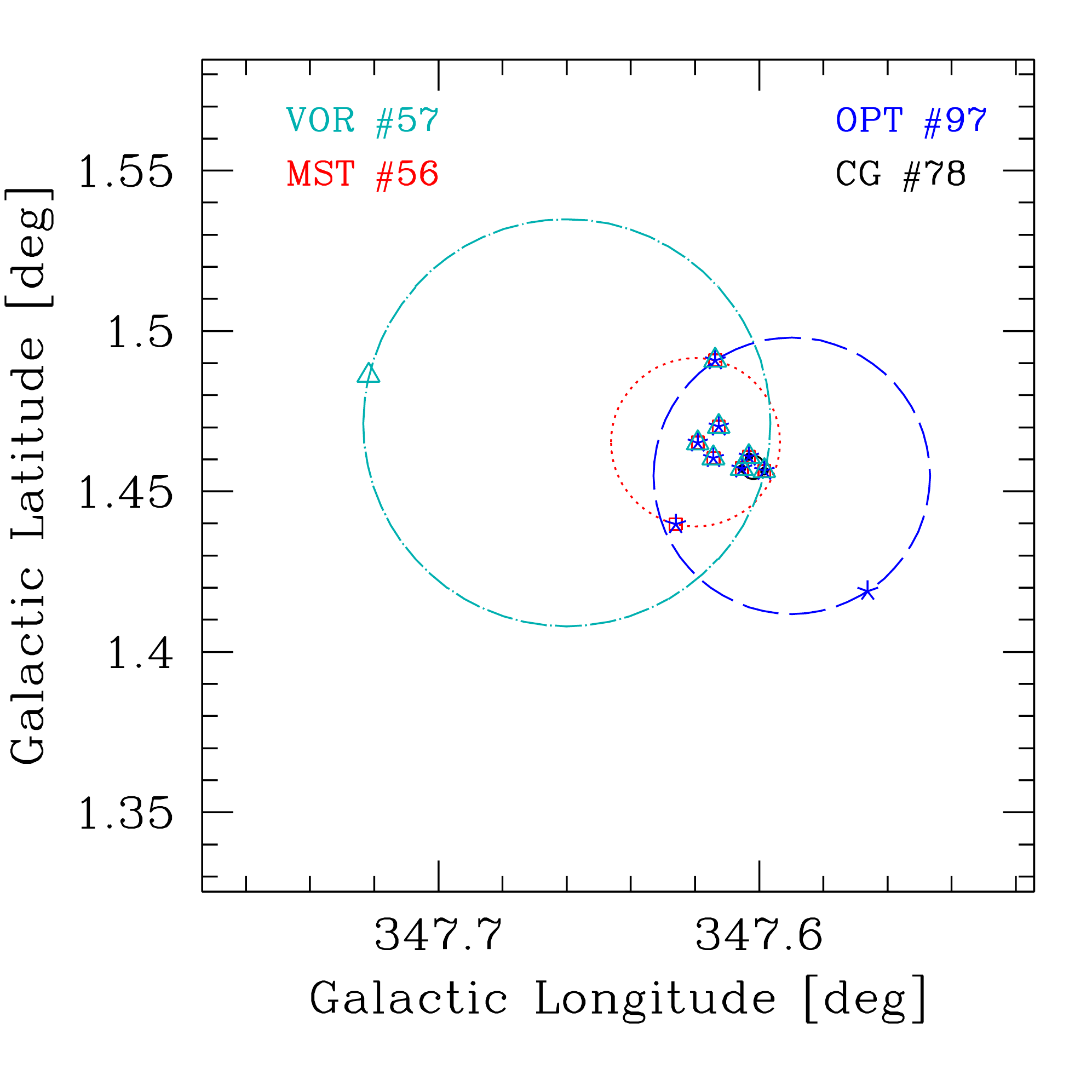}
    \includegraphics[width=0.24\textwidth]{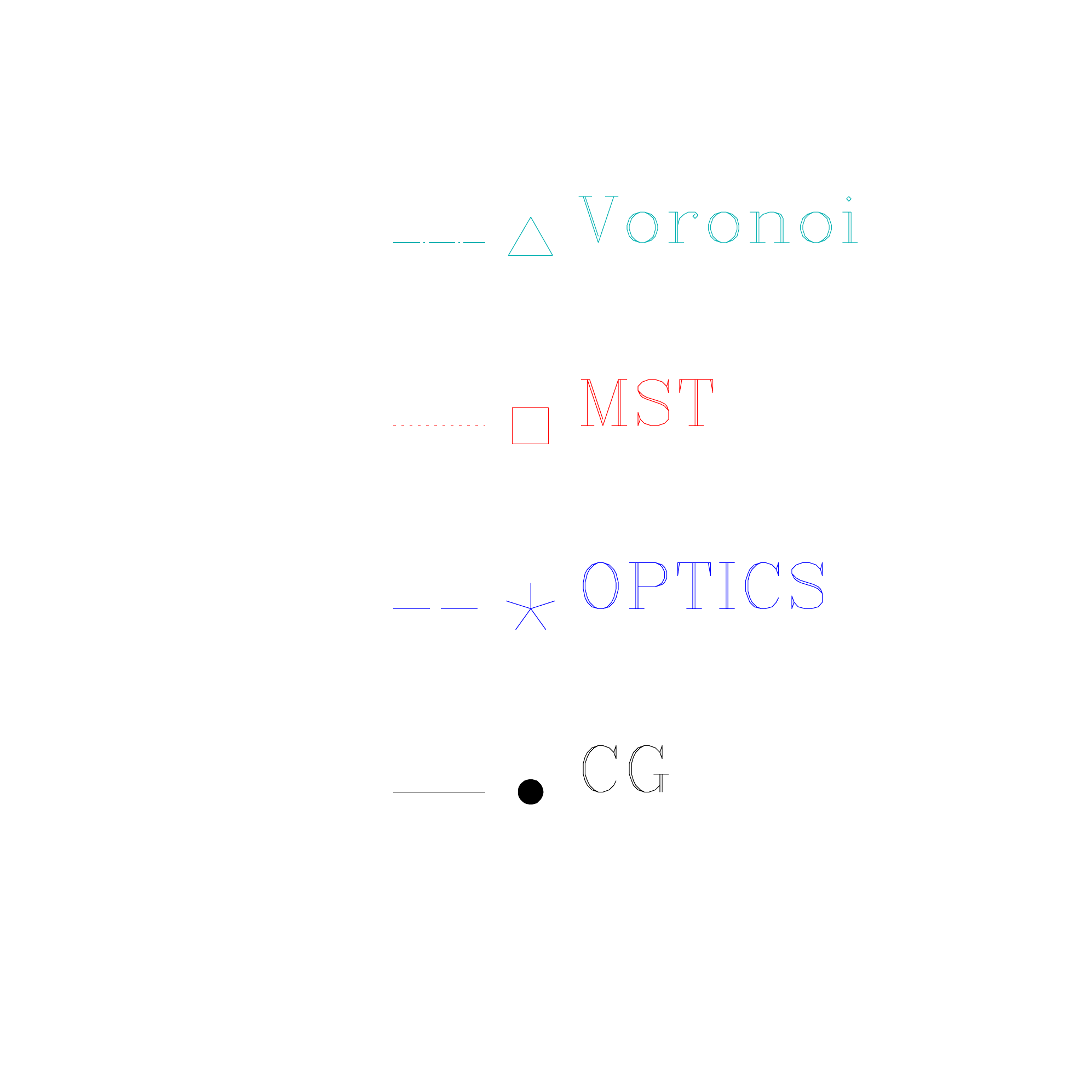}
    \caption{Overdensities identified by the four algorithms in the sky projection.  Circles represent the minimum circle that encloses the galaxy members.}
    \label{fig:cuadruple}
\end{figure*}


\bsp	
\label{lastpage}

\end{document}